\documentclass[12pt, letterpaper, oneside]{article}

\usepackage{graphicx}
\usepackage[body={6.6in, 9in}, right=1in, top=1in]{geometry}
\usepackage{amssymb}
\usepackage{amsmath} 
\usepackage{feynmp} 
\usepackage{subfigure}
\usepackage[linktoc=page,colorlinks=false,linkcolor=cyan,citecolor=cyan,urlcolor=blue]{hyperref}

\def\be{\begin{equation}}
\def\ee{\end{equation}}
\def\ba{\begin{eqnarray}}
\def\ea{\end{eqnarray}}	
\def\l{\left}
\def\r{\right}
\def\fr{\frac}
\def\la{\label}
\def\d{\partial}

\newcommand{\sfrac}[2]{{\textstyle\frac{#1}{#2}}}
\def\xdot{\dot{\vec x}}

\begin{document}

\begin{fmffile}{graphs}


\begin{center}
\Large{\textbf{Mutual Interactions of Phonons, Rotons, and Gravity}} \\[1cm]
\large{Alberto Nicolis$^{\rm a}$ and Riccardo Penco$^{\rm b}$}
\\[0.4cm]

\vspace{.2cm}
\small{\textit{$^{\rm a}$ Center for Theoretical Physics and Department of Physics, \\
  Columbia University, New York, NY 10027, USA}}

\vspace{.2cm}
\small{\textit{$^{\rm b}$ Center for Particle Cosmology, Department of Physics and Astronomy, \\
University of Pennsylvania, Philadelphia, Pennsylvania 19104, USA}}

\end{center}

\vspace{.2cm}


\begin{abstract}
We introduce an effective point-particle action for generic particles living in a zero-temperature superfluid. This action describes the motion of the particles in the medium at equilibrium as well as their couplings to sound waves and generic fluid flows. While we place the emphasis on elementary excitations such as phonons and rotons, our formalism applies also to macroscopic objects such as vortex rings and rigid bodies interacting with long-wavelength fluid modes. Within our approach, we reproduce phonon decay and phonon-phonon scattering as predicted using a purely field-theoretic description of phonons. We also correct classic results by Landau and Khalatnikov on roton-phonon scattering. Finally, we discuss how phonons and rotons couple to gravity, and show that the former tend to float while the latter tend to sink but with rather peculiar trajectories. Our formalism can be easily extended to include (general) relativistic effects and couplings to additional matter fields. As such, it can be relevant in contexts as diverse as neutron star physics and light dark matter detection.
\end{abstract}






\section{Introduction}

Neutron scattering experiments~\cite{Yarnell:1959zz} show that the spectrum of elementary excitations in superfluid helium-4 at very low temperatures looks schematically as in Figure \ref{fig:1}. A similar dispersion relation was also observed numerically~\cite{PhysRevLett.90.250403} as well as experimentally~\cite{chomaz2017} in trapped gases made of weakly interacting dipolar particles. There are two regions of momenta where the corresponding excitations are always kinematically stable\footnote{Depending on the precise shape of the dispersion curve, excitations in the region surrounding the local maximum could also be kinematically stable. Such excitations are usually referred to as \emph{maxons}. The formalism that we will here can be applied to maxons as well.}: one is around $p=0$, the other is around $p = p_*$.   Excitations in these regions are usually treated as different species of particles and are referred to as \emph{phonons} and \emph{rotons} respectively. Phenomenologically, their dispersion relations can be extracted by Taylor expanding the experimental dispersion curve around $p=0$ and $p = p_*$ to obtain
\begin{eqnarray} \la{pheno disp rel}
E_{\rm phonon} \simeq c_s p \; , \qquad \qquad \qquad E_{\rm roton} \simeq \Delta + \fr{(p - p_*)^2}{2 m_*}\; .
\end{eqnarray}
The precise values of the parameters $c_s, \Delta, p_*$, and $m_*$ depend on temperature and pressure (or, equivalently, temperature and chemical potential), but their orders of magnitude are correctly determined by dimensional analysis alone. For instance, the only relevant microscopic quantities in liquid helium-4 are the mass of the helium atom $m$, the Bohr radius $a$, and the typical interatomic distance, also of order $a$. In terms of these quantities, we have
\begin{eqnarray} \label{dim analysis scalings}
c_s \sim \fr{1}{ma} ,\qquad \qquad \Delta \sim \fr{1}{ma^2}, \qquad \qquad p_* \sim \fr{1}{a}, \qquad \qquad m_* \sim m \; .
\end{eqnarray}
(Throughout the paper we are working in units such that $\hbar = k_B = 1$. Moreover, for non-relativistic systems such as the one at hand, the speed of light $c$ cannot appear in our estimates.)

It is important to stress that phonons and rotons have a very different status. In fact, phonons are the Goldstone modes associated with the spontaneous breaking of the particle number $U(1)$ symmetry, which occurs in all superfluids. Specifically, calling $Q$ the generator of such a symmetry and $H_0$ the Hamiltonian, superfluids can be thought of as systems that spontaneously break both while preserving the combination
\be \label{tilde H}
 H = H_0 - \mu Q \; ,
\ee
which is the relevant generator of time translations at finite chemical potential $\mu$ \cite{Nicolis:2011pv}.
The existence and physical properties of phonons follow exclusively from symmetry principles~\cite{Endlich:2013spa} and, as a consequence, they belong in the spectrum of any superfluid. 
On the other hand, the existence of rotons is not enforced by any symmetries and, in fact, not all superfluids feature roton-type excitations.

\begin{figure}[t]
\begin{center}
\includegraphics[width=0.4\linewidth]{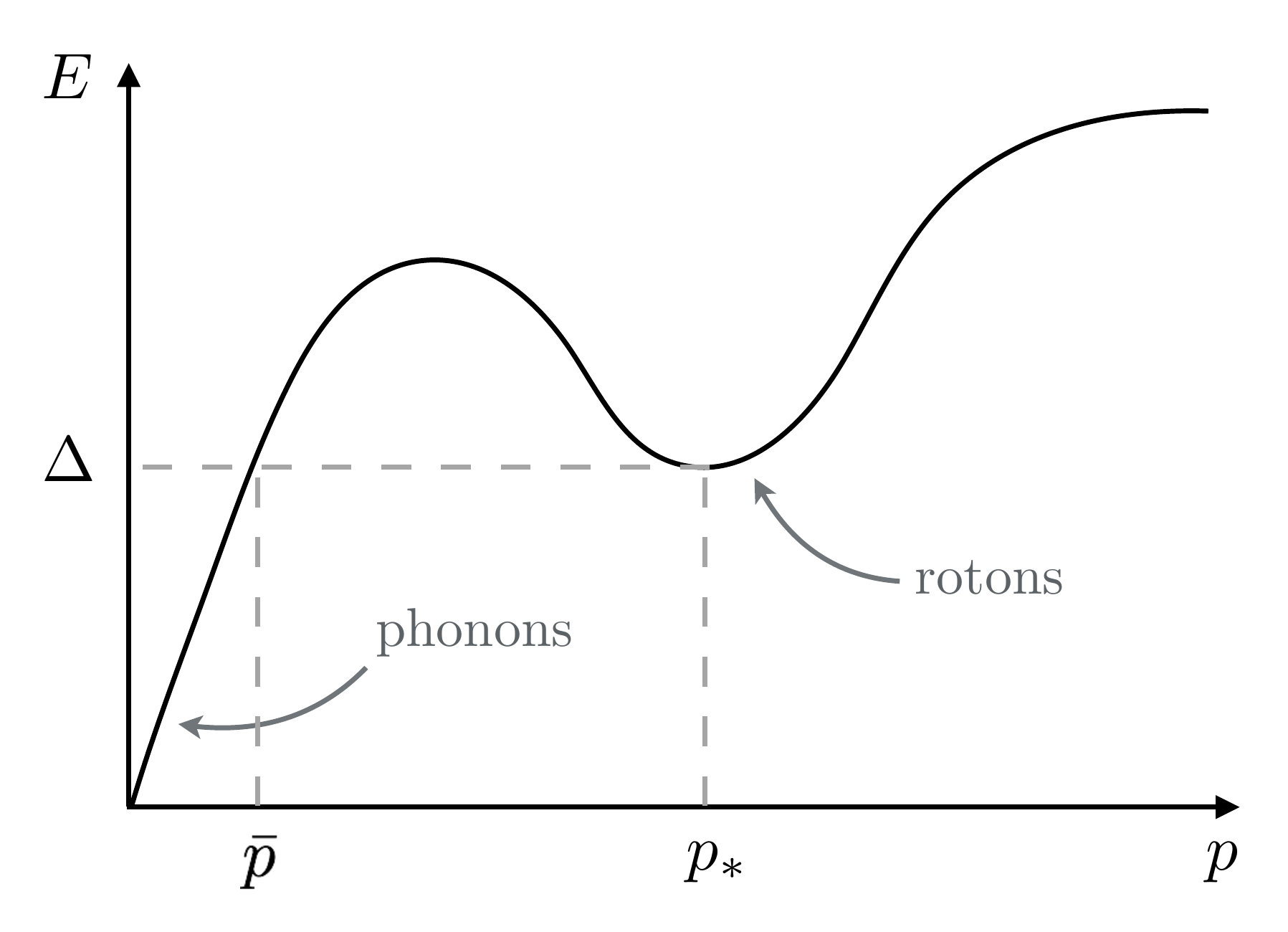}
\end{center}
\vspace{-0.5cm}
\caption{\small \it Prototypical phonon-roton dispersion curve.} \la{fig:1}
\end{figure}

This dichotomy also manifests itself in the field theory description of these two kinds of excitations. On the one hand, at energy scales much smaller than the roton gap $\Delta$, phonons are the only relevant degrees of freedom and admit a well-known effective field theory description based on the action~\cite{Greiter:1989qb}
\begin{eqnarray} \la{P(X) action}
S = \int d^4 x \, P(X), \qquad \qquad\qquad  X = {\mu}/{m} + \dot \pi - \tfrac{1}{2} (\vec \nabla \pi)^2.
\end{eqnarray}
Here $\pi(x)$ is the phonon field; it is a scalar under rotations, but transforms nontrivially under (Galilei or Lorentz) boosts. The quantity $X$ is the local value of the chemical potential per unit mass, which, in the presence of a nontrivial $\pi(x)$, differs from the equilibrium value $\mu/m$. Furthermore, the equation of motion for $\pi(x)$ coincides with the hydrodynamical equation for the superfluid, which allows one to infer the definition of other macroscopic quantities such as the pressure, mass density, and velocity fields: 
\begin{eqnarray} \label{p rho u}
p = P, \qquad \qquad \quad \rho  = \fr{d P}{d X} , \qquad \qquad \quad \vec u = - \vec \nabla \pi.
\end{eqnarray}
Different choices for the function $P(X)$ correspond to different equations of state for the superfluid. Moreover, by expanding the action (\ref{P(X) action}) up to quadratic order in $\pi$, one finds that the speed of sound is indeed equal to $dp / d\rho$, as expected for a compressional wave. Since in the following we will use `$p$' also to denote the momentum of our excitations, to avoid confusion from now on we will denote the pressure simply by `$P \,$'.

On the other hand, it is not obvious how to extend the action (\ref{P(X) action}) in order to include rotons in the effective field theory description. Perhaps the main difficulty in achieving this is that phonons are no longer meaningful degrees of freedom at the typical roton momentum scales, since at such high scales an infinite number of higher derivative corrections to \eqref{P(X) action} become important. Nevertheless, unconcerned by these theoretical challenges, rotons play a crucial role in determining low-temperature macroscopic quantities that can be measured in the lab, such as the specific heat and the viscosity coefficients~\cite{landau1949theory, landau1949theory2}. Notice that this is true even at temperatures as low as 1 K, even though the roton gap is $\Delta \sim 10$ K. This is because, compared to gapped excitations at zero momentum, the roton contribution to thermodynamic quantities has an enhanced phase space, due to their large typical momentum. It is therefore important to have both phonons and rotons under full theoretical control.

In this paper we are going to propose a unifying EFT framework to describe mutual interactions of phonons and rotons. Since rotons are stable at low energies, we are going to treat them as ``heavy'', point-like objects that can interact either with soft phonons or with each other by exchanging virtual phonons. Our formalism is heavily inspired by recent developments concerning vortex lines in relativistic fluids and superfluids~\cite{Endlich:2013dma,Horn:2015zna,Esposito:2017xzg,Mitsou-et-al-2016} as well as by the NRGR approach~\cite{Goldberger:2004jt,Goldberger:2007hy} to the dynamics of non-relativistic extended objects coupled to gravity.

Since phonons and rotons lie on the same dispersion curve, they  are usually said to to be the same kind of excitation, just with different momenta. It is not obvious what operational meaning to attach to that statement. In systems with boost invariance, either of the Galilei or the Lorentz type, we have a symmetry that relates identical particles with different momenta: a very energetic electron is the same as an electron at rest, seen from a highly boosted reference frame. Here instead, the surrounding medium breaks boosts (spontaneously), and so there is no symmetry that relates particles with different momenta. The most precise meaning we can find for the above statement is: {\em (i)} Phonons and rotons have the same quantum numbers, that is, they transform in the same way under the symmetries, and {\em (ii)} in the energy vs. momentum plane for states with those quantum numbers, at low enough energies there is only one line occupied by single particle states (for any given momentum, at higher energies there will be a continuum occupied by multi-particle states).

From this viewpoint, the qualitative difference between rotons and phonons is no more dramatic than that between two phonons of different momenta. In the following we will invest in this idea, and argue that from the standpoint of an effective theory for a point-particle interacting with long wavelength bulk modes (i.e.~soft phonons or macroscopic fluid flows), phonons, rotons, and even macroscopic objects such as vortex rings and rigid bodies can all be described by the same general Lagrangian, expanded about different values of the point particle's velocity and with different parameters.

As a check of our formalism, we will compute the rates for certain phonon processes in kinematical regimes that are also amenable to a more standard effective field theory analysis, and we find perfect agreement between the two approaches. Encouraged by these results, we will compute 
the cross section for roton-phonon scattering, a process originally considered by Landau and Khalatnikov in~\cite{landau1949theory}. We find new interaction terms that had been overlooked in their computation.

Finally, in the point-particle limit it becomes straightforward to discuss an aspect rarely considered in the condensed matter literature, namely the coupling of collective excitations to gravity. In this respect, phonons resemble photons to the extent that their effective gravitational mass is proportional to their momentum, but differ in a very important way: they have a negative effective gravitational mass, so their trajectories bend upwards rather than downwards in a gravitational field. This result can also be understood in more conventional terms using the language of wave mechanics. However, our results concerning rotons are novel: we find that rotons in superfluid helium-4 tend to sink, although they do so by following very peculiar trajectories. While the individual trajectories are unlikely to be directly observable, it is at least plausible that they might lead to measurable effects in the aggregate, especially when compared with the qualitatively different behavior of phonons. For instance, at nonzero temperature the thermal distributions of phonons and rotons will depend on height in different ways.

\section{The effective point-particle theory}\label{effective}

Consider one of the excitations of Fig.~\ref{fig:1}, propagating in a homogeneous superfluid at rest. According to standard representation theory, single particle states have to fall into irreducible representations of the unbroken symmetries. Our medium breaks spontaneously the original internal $U(1)$ symmetry and boosts, but preserves spatial translations, time translations (modified as in \eqref{tilde H}), and rotations. For given momentum $\vec p$, as usual, one looks for irreducible representations of the little group of $\vec p$, which in this case are nothing but rotations around $\vec p$. The associated quantum number is the particle's helicity, and we already know that phonons are zero-helicity particles, since they are the quanta of longitudinal compressional waves. Assuming that all excitations of fig.~\ref{fig:1} have the same quantum numbers as phonons (apart from $\vec p\,$), we thus see that the single-particle states we are interested in are labelled just by $\vec p$, and have no other degrees of freedom.

The curve in Fig.~\ref{fig:1} gives the Hamiltonian as a function of $\vec p$,
\be
E = H (|\vec p \, |) \; .
\ee
In the point-particle limit, that is if we consider a small wave packet localized around $\vec x(t)$, the associated classical Lagrangian is just the Legendre transform of this:
\be
L_{\rm p.p.} = \vec p \cdot \dot{\vec x} - H(|\vec p \,|) = f(|\dot{\vec x}|)\; , \qquad {\dot{\vec x}} = \frac{\partial H}{\partial \vec p} \; .
\ee
Such a Lagrangian is clearly invariant under translations and rotations. It is not invariant under Galilei boosts, but that has to be expected since the surrounding medium breaks boosts. In fact, if we now consider turning on long-wavelength perturbations in the surrounding medium, Galilei invariance forces $\dot {\vec x}$ to always appear in the combination $\dot {\vec x} - \vec u$, where $\vec u$ is the local fluid velocity at the particle's position. Moreover, the parameters that define the function $L_{\rm p.p.}$, who are in one-to-one correspondence with those that define the curve $E=H(|\vec p|)$ in fig.~\ref{fig:1}, can now depend on the local invariants one can construct using the bulk degrees of freedom. To lowest order in the bulk modes' derivatives, the only such invariant is $X$, defined in Eq.~(\ref{P(X) action}). In the presence of perturbations $\pi(x)$, the most general point-particle action for helicity-zero particles thus is
\be \label{Spp}
S_{\rm p.p.} [\vec x, \pi]= \int dt \,  f(|\dot {\vec x} - \vec u|, X) \; .
\ee

Notice that, in principle, we could add to this action the standard Galilean kinetic energy:
\be \label{v^2}
\int dt \, \sfrac12 M_i \, \dot {\vec x} \,^2 \; ,
\ee
with constant inertial mass $M_i$. This is invariant under Galilean boosts only up to total derivatives, which is why it is not contemplated by the form \eqref{Spp}. However, recalling that $\vec u = -\vec \nabla \pi$ and $X =\mu/m +  \dot \pi - \frac12 (\vec \nabla \pi)^2$, after straightforward manipulations we get
\be \label{manipulate}
\sfrac12 M_i \, \dot {\vec x} \,^2 = \sfrac12 M_i |\dot {\vec x} - \vec u |^2 + M_i(X-\mu/m) + \sfrac{d}{dt} \big(M_i \pi\big) \; ,
\ee
which is in fact of the form \eqref{Spp} up to a total derivative term. We can thus restrict ourselves to the original action \eqref{Spp}.

This action includes the most general interactions of our particle with long-wavelength bulk modes, such as fluid flows, pressure gradients, or soft phonons. We should keep in mind though that, according to standard effective field theory logic, the action \eqref{Spp} is just the leading order one in a derivative expansion. Subleading corrections will  involve:
\begin{itemize} 
\item Higher spatial derivatives acting on the bulk fields: these are suppressed by coefficients that scale like the appropriate power of the typical size of our particle. For small material objects or classical fluid configurations (such as a vortex ring), this is just their size; for quantum excitations such as phonons and rotons, this is the generalized de Broglie wavelength, $1/p$.
\item Higher time derivatives acting on the bulk fields or on $\vec x(t)$: these are suppressed by coefficients that scale like the appropriate power of the typical internal timescale for our particle. This be could the period of the slowest normal mode if our particle has any, or just the sound-crossing time---the typical size defined above over $c_s$---, whichever is longer.
\end{itemize}
For all the computations that follow, the lowest-order action above will suffice.

The symmetries of the system do not constrain the functional dependence of the function $f$ in \eqref{Spp} any further. The actual function has to be determined from experimental data. However, it is interesting that, at this order, for phonons and rotons all the necessary information is contained in the $H(|\vec  p|)$ of fig.~\ref{fig:1} and in  its dependence on any thermodynamic quantity {\em at equilibrium}, such as the pressure or the chemical potential:
At  $X= \mu/m$, $f$ as a function of $|\dot {\vec x} - \vec u|$ is just the Legendre transform of $H(|\vec p|)$; the $X$ dependence instead can be inferred by looking at how the parameters in $H(|\vec p|)$ vary with pressure or chemical potential. We will see explicitly how this works for phonons, rotons, and vortex rings. In particular, for phonons and vortex rings, the functional form of $f$ is uniquely determined by the superfluid's equation of state.

Because of certain technical subtleties that we will encounter in taking the Legendre transform of $H$ in the phonon and roton limits, it is useful to keep in mind that an action can always also be interpreted as a functional of the $q$'s and the $p$'s separately,
\be \label{Sqp}
S[q,p] \equiv \int dt \, p_\alpha \dot q^\alpha - H(q,p) \; .
\ee
The variational problem with arbitrary $\delta q $ and $\delta p$ (with fixed boundary conditions for $q$) yields Hamilton's equations in this case. {\em If} the 
$\dot q = \partial H/ \partial p$ equations can be solved for all the $p$'s, then one can plug back the solutions into the action and obtain the usual Lagrangian formulation, with a variational principle for $S[q]$. If instead some of the $p$'s cannot be integrated out, one can keep them explicitly in the action as in \eqref{Sqp}. In fact, from a quantum mechanical viewpoint the mixed $q, p$ formulation in \eqref{Sqp} is the fundamental one, since that's what appears in the path integral starting from the canonical formalism.

\subsection{Phonons}

Phonons correspond to the low-momentum region in Fig.~\eqref{fig:1}. For the moment, let's set to zero the perturbations in the surrounding fluid. The phonon Hamiltonian thus is
\be
H = c_s |\vec p \,| + \dots
\ee
where $c_s$ is the sound speed, and the dots stand for higher powers of $|\vec p \, |$.
To take the Legendre transform, we first need to solve 
\be \label{full eq}
\dot{\vec x} = \frac{\d H}{\d \vec p} \simeq c_s  \hat p
\ee
for $\vec p$. However, at this order this equation does not involve the 
magnitude $p$, and thus cannot be solved to eliminate $\vec p$ completely.  
As recalled above, we can still define the action as usual, but now we should treat $p$ as an independent variable.
So, starting from the full action functional for $p$, $\hat p$, and $\vec x$,
\begin{eqnarray} \la{L phonons} 
S \equiv \int dt \, ( \dot {\vec x}\cdot \vec  p - H ) \simeq  \int dt \, p \, ( \dot {\vec x}\cdot \hat  p - c_s ) \; ,
\end{eqnarray}
we only impose the Hamilton equation associated with varying $\hat p$, which,  recalling that $\hat p $ is constrained to have unit norm, reads
\be \label{p hat solution}
(\delta_{ij} - \hat p_i \hat p_j) \dot x_j =0 \qquad \Rightarrow \qquad \hat p = \frac{ \xdot}{| \xdot|} \, . 
\ee
Notice that in principle the equation above doesn't determine the sign of $\hat p$. However, it is easy to check that the solution with the opposite sign, i.e. $\hat p = - \xdot / | \xdot|$, would lead to a dynamical system that doesn't admit any solution for $c_s >0$. This can already be seen at the level of eq.~\eqref{full eq}. Plugging therefore the solution \eqref{p hat solution} into the action, we finally get an action functional for $p$ and $\vec x$ only:
\begin{eqnarray} \la{L phonons} 
S[\vec x, p] \simeq \int dt \,  p \, \big(|\dot{\vec x}| - c_s \big) \; .
\end{eqnarray}

Alternatively, we could have kept the first correction to the phonon Hamiltonian,
\begin{eqnarray}
H = c_s \, |\vec p \,|  + \fr{c_s}{\Lambda^2} \fr{|\vec p \, |^3}{3} + \dots,
\end{eqnarray}
where  $\Lambda$ is some large momentum scale of order $1/a$, which eventually we would like to send to infinity (in the sense that $p \ll \Lambda$).  Such a cubic term is indeed needed to describe the phonon spectrum beyond leading order, and it makes phonon decay a kinematically allowed process in helium-4~\cite{landau9}.
The velocity now reads
\begin{eqnarray} \la{v of phonon}
\xdot =  \fr{\d H}{\d \vec p} \simeq c_s \hat p \l[1 + \fr{p^2}{\Lambda^2}   \r] \; ,
\end{eqnarray}
which allows us to eliminate $\vec p$ completely:
\be
\hat p \parallel \xdot \; , \qquad p = \Lambda \big(|\xdot|/c_s - 1 \big)^{1/2} \; .
\ee
The resulting action is:
\begin{eqnarray} \la{L phonons Lambda}
S[\vec x] = \int dt \,  \sfrac{2}{3} \, {\Lambda c_s} \big(|\xdot |/c_s - 1 \big)^{3/2}.
\end{eqnarray}

At this stage, the limit $\Lambda \to \infty$ is clearly not well defined. This is of course just a reflection of the fact that Eq.~(\ref{v of phonon}) cannot be solved for $p$ in this limit. An analogous situation occurs for an ordinary relativistic point particle, whose Lagrangian does not admit a straightforward massless limit. In that case, the $m \to 0$ limit can be taken only after introducing an auxiliary variable playing the role of an ``einbein''~\cite{Polchinski:1998rq}. A similar procedure allows one to go from the Nambu-Goto action for a relativistic string to the Polyakov action by introducing a dynamical metric on the worldsheet---although in that context this step is taken to simplify the quantization procedure rather than to take a zero-tension limit.

We can follow a similar approach, and rewrite the Lagrangian (\ref{L phonons Lambda}) by introducing an auxiliary variable, which with the benefit of hindsight we are going to denote as $p$:
\begin{eqnarray} \la{L phonons auxiliary variable}
S[\vec x, p] = \int dt \, p \, \big(|\xdot| - c_s \big) - \fr{c_s p^3}{3 \Lambda^2} \; .
\end{eqnarray}
It is easy to check that after integrating out $p$ one indeed recovers  the Lagrangian (\ref{L phonons Lambda}). Written in this form, though, $S$ remains well defined in the limit $\Lambda \to \infty$, where it reduces to our original action \eqref{L phonons}.

We can now introduce long wavelength bulk modes $\pi(x)$ according to the general prescription outlined above. We get
\begin{eqnarray} \la{Sphonon}
S_{\rm phonon}[\vec x, p, \pi] = \int dt \, p  \big[ \, | \xdot - \vec u| - c_s (X) \big] \; .
\end{eqnarray}
Recall that $c_s^2 = dP/d \rho$, and that $X$ can be interpreted as the local chemical potential per unit mass. The function $c_s(X)$ is thus uniquely determined by the superfluid's equation of state at equilibrium, i.e.~by how  $P $ and $\rho$ depend on the chemical potential. 

We can use the action above to describe the interactions of our phonon with much softer ones. All the couplings  follow  from expanding this action in powers of the bulk phonon field $\pi(x)$. Of course, interactions that only involve phonons can also be described using the effective theory (\ref{P(X) action}). The point-like action (\ref{Sphonon}) offers an alternative viewpoint which, as we will see in Sec. \ref{scattering}, is completely equivalent to the field theory approach.

\subsection{Rotons}

Rotons correspond to excitations close to the minimum in Fig.~\ref{fig:1}. They are stable for kinematical reasons, as shown in Appendix \ref{appa}. In the absence of external perturbations, their Hamiltonian reads
\be
H = \Delta + \fr{(|\vec p \,| - p_*)^2}{2 m_*} + \dots\; ,
\ee
where the dots stand for higher powers of $(|\vec p \, | - p_*)$. The velocity/momentum relationship now is
\be \label{velocity roton}
\xdot = \frac{\d H}{\d \vec p} \simeq \hat p \frac{(p-p_*)}{m_*} \; ,
\ee
which can be inverted, but yields two branches of solutions for the momentum:
\begin{align}
p > p_*: & \qquad \hat p \parallel \xdot \; , \qquad p = p_* + m_* |\xdot| \\
p < p_*: & \qquad \hat p \parallel - \xdot \; , \qquad p = p_* - m_* |\xdot|
\end{align}
Correspondigly, there are two branches for the resulting action, depending on whether the roton is to the right ($R$) or to the left ($L$) of the minimum:
\be
S^{R,L}  [\vec x \, ] \simeq \int dt \, \big[ -\Delta \pm p_* |\xdot| + \sfrac12 m_*{ |\xdot|^2} \big] \; .
\ee
Clearly, this action is singular right at the minimum, where $\xdot = 0$, which is also related to the fact that eq.~\eqref{velocity roton} cannot be solved for the direction $\hat p$ for zero velocity. So, for computations  that require regularity at $\xdot=0$, one can refrain from integrating out $\hat p$, in which case both $p$ and the action are single-valued and regular at any  $\xdot$:
\be
p = p_* + m_* (\xdot \cdot \hat p) \; , \qquad  S [\vec x, \hat p ] \simeq \int dt \, \big[ -\Delta + p_* (\xdot \cdot \hat p) + \sfrac12 m_*{ (\xdot \cdot \hat p)^2} \big] 
\ee
Introducing  bulk perturbations as above, we finally get
\be \label{effective action for rotons1}
S_{\rm roton}^{R,L}  [\vec x , \pi] \simeq \int dt \, \big[ -\Delta(X) \pm p_*(X) |\xdot - \vec u| + \sfrac12 m_*(X) { |\xdot - \vec u|^2} \big] \; ,
\ee 
or, equivalently,
\be \label{effective action for rotons2}
S_{\rm roton}  [\vec x , \hat p, \pi] \simeq \int dt \, \big[ -\Delta(X) + p_*(X) (\xdot -\vec u)\cdot \hat p + \sfrac12 m_* (X) { \big((\xdot -\vec u)\cdot \hat p \big)^2} \big]  \; .
\ee

Like for phonons above, the actions (\ref{effective action for rotons1}, \ref{effective action for rotons2}) describe all  possible low-energy interactions of a roton with long wavelength bulk modes. These interactions can be deduced by expanding the action in powers of the bulk phonon field $\pi(x)$. The associated coupling constants  involve $X$-derivatives of the parameters $\Delta$, $p_*$, and $m_*$, which can be traded for derivatives with respect to pressure, which for helium-4  have been measured experimentally~\cite{woods1973structure}.

%

\subsection{Vortex rings}
We can apply the same formalism to describe macroscopic circular vortex rings and their interactions with long wavelength bulk modes.
Clearly, in no sense is a macroscopic vortex ring an elementary  excitation. However, from the viewpoint of our symmetry considerations above, a circular ring has the same transformation properties as a phonon or a roton: definite momentum, and  zero helicity. 
It should thus be possible to describe it in the point particle limit using the general formalism developed above.  Let's see how this works.

Starting from the classical action for a vortex line in an unperturbed fluid, one can parametrize a circular vortex ring by its center's position $\vec x(t)$, its orientation---the normal unit vector $\hat n(t)$---and its radius $R(t)$, and end up with an effective action for these degrees of freedom only \cite{Horn:2015zna}
\be \label{Sring}
S  [\vec x, \hat n, R ]= \int dt \l[ \pi \bar \rho \, \Gamma \,  R^2 \big(\hat n \cdot {\dot {\vec x}}\big) -2 \pi R \, T(1/R) \r] \; ,
\ee
where $\Gamma$  is the vortex line's circulation, $\bar \rho$ the background mass density, and $T$ the vortex line's energy per unit length (i.e., its tension), which runs logarithmically with momentum scale \cite{Horn:2015zna}:
\be
T(1/R) = \frac{\bar \rho \, \Gamma^2}{4 \pi} \log(R  p_0) \; ,
\ee 
where $p_0$ is a UV momentum scale, typically of order of the string's inverse thickness $\sim 1/a$, but logically separate from it.\footnote{If needed, the exact value of $p_0$ has to be determined from experiment. However, for rings much bigger than the string's thickness, the log is large and one can safely replace $p_0$ with $1/a$.}

However, we can see right away that $R$ and $\hat n$ appear in the action without time derivatives, and  can thus be integrated out. In fact, it's clear that together they play the role of the conjugate momentum to $\vec x$:
\be
\vec p = \frac{\d L}{\d \xdot} = \pi  \bar \rho \, \Gamma  \cdot  R^2 \hat n \; ,
\ee
so that the action above is really a mixed $q,p$ action of the form \eqref{Sqp}:
\be
S  [\vec x, \vec p \,  ] = \int dt \, \vec p \cdot \xdot - H(|\vec p|) \; , 
\ee
with associated Hamiltonian
\be
H(p) = \sqrt{\frac{\bar \rho \, \Gamma^3}{16\pi } \cdot p} \, \log\Big( \frac{p_0
^2  \, p }{\pi \bar \rho \, \Gamma} \Big) \; .
\ee

Solving $\xdot = {\d H}/{\d \vec p}$ for $\vec p$ we get 
\begin{align}
\hat p \parallel \xdot \; , \qquad p \simeq \frac{\bar \rho \, \Gamma^3}{16\pi |\xdot|^2} \log^2  \frac{\Gamma p_0}{8 \pi |\xdot|} 
\end{align}
(we are assuming that the logs are large and positive, so that one can invert the relation $y = \frac{\log z}{z}$ approximately by $z \simeq \frac{\log{1/y}}{y}$.) 
Plugging back into the action, we get the desired action for $\vec x(t)$ only
\begin{align}
S[\vec x \, ] \simeq - \int dt \, \frac{\bar \rho \, \Gamma^3 }{16\pi } \frac{1}{|\dot {\vec x}|} \log^2 \frac{\Gamma p_0}{8 \pi  |\dot {\vec x}|} \; . 
\end{align}

Using the same logic as above, we can now introduce long wavelength bulk modes simply by replacing $\xdot$ with the relative velocity $\xdot - \vec u$, and the background quantities $\bar \rho$ and $p_0$ with their $X$-dependent counterparts. Notice that the  circulation $\Gamma$  does not depend on bulk quantities for a non-relativistic superfluid (it is in fact quantized in units of $1/m$). Notice also that, within our large-log approximation, the $X$-dependence of $p_0$ can be safely ignored. We thus get
\begin{align}
S_{\rm ring}[\vec x, \pi]  \simeq - \int dt \, \frac{\rho(X) \Gamma^3}{16\pi} \frac{1}{|\xdot - \vec u|} \log^2 \Big( \frac{\Gamma p_0}{8 \pi |\xdot - \vec u|} \Big) \; .
\end{align}
We thus see that the interactions of a circular vortex ring with much longer bulk modes are completely constrained, in the sense that they are uniquely determined by the symmetries and by the superfluid's equation of state, via the function $\rho( X)$.

A macroscopic vortex ring with $R \gg a$ has a typical speed much lower than that of sound: up to logs, we have $v \sim c_s (a/R)$. However, the zero-velocity limit is singular, as clear from the action above, since it corresponds to infinitely large rings. We can thus think of rotons and vortex rings as qualitatively different low-velocity point-particles. Their difference is clearer in the Hamiltonian formulation: rotons have typical momenta of order $p_* \sim 1/a$ and typical energies of order $\Delta \sim c_s/a$, while vortex rings have much bigger momenta and energies: up to logs,
\be
p \sim \frac{\rho \Gamma^3}{v^2} \gg 1/a \; , \qquad H(p) \sim \sqrt{\rho \Gamma^3 p} \gg c_s/a \; .
\ee
From this viewpoint, macroscopic vortex rings with, say, one quantum of circulation can be thought of as occupying the far right of the energy spectrum in Fig.~\ref{fig:1}, with an energy scaling as $\sqrt p \log p$. (This statement however should be taken with a grain of salt, since experimentally one finds that the thin curve of fig.~\ref{fig:1} gets in fact significantly smeared out on the right by a two-roton state continuum.)

\subsection{Objects}
Finally, we can also use our formalism to describe how an ordinary material object (or an ordinary massive particle) interacts with long-wavelength modes in the surrounding superfluid. We will publish elsewhere a more complete analysis that takes into account possible anisotropies in the object's shape as well as the effects of considering an ordinary fluid rather that a superfluid. Here instead we just want to mention some basic facts.

What is an ``ordinary object"? To some extent, the answer is a matter of definition. However, we can notice that phonons, rotons, and vortex rings all correspond to somewhat peculiar limits of our general action \eqref{Spp}. Phonons have small momenta ($p \ll p_*$) but large velocities ($v \simeq c_s$), rotons and vortex rings have small velocities ($v \ll c_s$) but large momenta ($p \simeq p_*$ and $p \gg p_*$.) 
We can thus define an ordinary object in our point-particle limit as a particle described by an action of the general form \eqref{Spp}, but with an ordinary $v \to 0$, $p \to 0$ limit: at least at low speeds, we want $p \propto v$. We thus have (in the presence of generic perturbations in the superfluid)
\be \label{Sobj}
S_{\rm obj} = \int dt \big[- E_0(X) + \sfrac12 M_{\rm eff}(X) { |\xdot - \vec u|^2} + \dots \big] \; ,
\ee
where the dots stand for higher powers  of $|\xdot - \vec u|$, which we expect to be suppressed by inverse powers of $c_s$. Our definition of an object tells us nothing about the structure of the action at high (relative) speeds, which suggests that there might not be any fundamental difference between ordinary objects and more general excitations at high speeds.

The $E_0$  term in the action above measures the rest energy of the object. It can depend on the local value of $X$, which is related to the local pressure. This is obvious for a compressible object---such as a balloon---but in fact the $X$-dependence of $E_0$ is associated which much  more general effects, as we will see in Sec.~\ref{gravity}. The coefficient of the kinetic energy $M_{\rm eff}$ also depends on the local value of $X$, and in general can be very different from the mass of the object in empty space: on general grounds, one expects  the interactions between the object and the surrounding superfluid to yield contributions to $M_{\rm eff}$ of order of the mass of the fluid displaced
(see e.g.~\cite{landau:1987bo}, \S 11, Problem 1). In our formalism, these come from a classical self-energy diagram in which the object exchanges a $\pi$ field with itself.

\section{A check: phonon processes}\label{scattering}
As a check of our point-particle formalism, we will now compute two processes involving phonons, and show that they match the results obtained from the effective field theory described by \eqref{P(X) action}.
There is in fact a nontrivial overlap of the regimes of validity of the two effective theories: Our point-particle theory is valid for a phonon of momentum $p$ interacting with much longer bulk modes, such as phonons with typical momenta $k \ll p$; the $P(X)$ effective field theory is valid for any number of low-energy phonons, with momenta much smaller than the UV cutoff $p_* \sim 1/a$. If we consider processes with one incoming and one outgoing phonon with typical momenta $p \ll p_*$,  and any number of much softer incoming or outgoing phonons, we should be allowed to use either theory. We will call `hard' the former phonons, and `soft' the latter.

As a first application, consider the decay of a hard phonon into a hard one and a soft one. This is nothing but Cherenkov sound emission by the hard phonon. It is kinematically possible in helium 4, because the phonons' dispersion law is $E = c_s p + \alpha p^3 +\dots$, with positive $\alpha$, and so hard phonons have faster propagation speeds than soft ones. In both descriptions, there is only one diagram contributing to this process to leading order.

In the $P(X)$ theory, we need the expansion of the action up to cubic order in $\pi$:
\begin{align}
S  = \int d^4 x P(X) \to \int d^4 x \, \frac{\rho}{c_s^2} \bigg\{ \frac12 \left[\dot \pi^2 - c_s^2 (\vec \nabla \pi)^2 \right]  + \frac{g_3}{3! c_s^2} \, \dot \pi^3 - \frac{1}{2} \, \dot \pi (\vec \nabla \pi)^2 + \dots \bigg\} \; , \label{P(X) expanded}
\end{align}
where $g_3$ and all the $g_n$'s below are equation of state-dependent dimensionless coupling constants defined as
\be
g_n = c_s^{2(n-2)} P^{(n)}/P'' \; ,
\ee
evaluated at the $X = \mu/m$ equilibrium value, and
 $\rho$ and $c_s$ are also equilbrium values.
Denoting by $\vec k$  and $\vec p$ the momenta of the outgoing soft phonon and of the incoming hard one, and using the fact that $k \ll p$, we find that the amplitude~is\footnote{Notice that each external phonon line comes with a factor of $c_s / \sqrt{\rho}$ because our  $\pi$ field is not canonically normalized.}
\be \label{1to2 P(X)}
i {\cal M} \simeq \!\!\!\!\!\!\!\!\! \parbox{20mm}{
\begin{fmfgraph*}(50,40) 
\fmftop{i}
\fmfbottom{o}
\fmfdot{v}
\fmf{dashes}{i,v,o}
\fmffreeze
\fmfright{or}
\fmf{dashes}{v,or}
\end{fmfgraph*}
} \!\!  \simeq 
 \frac{2 c_s^2 }{\sqrt \rho} \, p^2 k \,  \big[ \cos\theta +\sfrac12 (1-g_3) \big] \; ,
\ee
with $\theta$ the angle between the two momenta. The associated decay rate is
\begin{align} \label{emission}
d \Gamma & = \frac{1}{2E} |{\cal M}|^2 d\Pi_f \\
&= \frac{c_s^2 p^2 k^2 }{\rho} \big[ \cos\theta +\sfrac12 (1-g_3) \big]^2 \times \frac{d^3 k}{(2\pi)^3 \, 2c_s k} \times (2\pi) \delta\big(E-(E' + \omega) \big) \; ,
\end{align}
where in the last line we have used the momentum-conserving delta  to integrate over the final hard phonon's momentum.
The energy-conserving delta can be simplified in the small $k$ limit: from $E' = E(|\vec p - \vec k|)$, we get
\be
\delta(E - (E'+\omega)) \simeq \delta (\omega - \vec v_0 \cdot \vec k ) \; ,
\ee
where
\be
\vec v_0 = \frac {\d E}{\d \vec p}
\ee
is the hard phonon's group  velocity. This gives the usual Mach-cone condition for Cherenkov radiation,
\be
\cos \theta = c_s/v_0 \; .
\ee
Integrating over the solid angle we finally get the emission rate per unit frequency:
\be
\frac{d \Gamma}{d \omega} = \frac{p^2}{16\pi \rho c_s^3} \, ( 3-g_3)^2\,  \omega^2 \; ,
\ee
where we took the limit $v_0 \to c_s^+$.
Notice that the final state phase space is nonzero even in this limit, that is, even if one neglects the higher order corrections to the dispersion law that make the decay possible in the first place.

Consider now the same process in the point-particle effective theory. We need to expand the action \eqref{Sphonon} to first order in the external $\pi$,
\be
S_{\rm phonon} \supset \int dt \, p\big( \, \hat v_0 \cdot \vec \nabla \pi - c_s' \dot \pi \big) \; ,
\ee
where $c_s'$ is the $X$-derivative of $c_s$. Treating the hard phonon as a source and neglecting its change in momentum from initial to final state, the amplitude for emission of a soft phonon is simply
\be
i {\cal M} \simeq \!\!\!\!\!\!\!\!\! \parbox{20mm}{
\begin{fmfgraph*}(50,40) 
\fmftop{i}
\fmfbottom{o}
\fmfdot{v}
\fmf{dbl_plain}{i,v,o}
\fmffreeze
\fmfright{or}
\fmf{dashes}{v,or}
\end{fmfgraph*}
} \!\! \simeq 
 \frac{c_s p k }{\sqrt \rho} (\cos\theta + c_s' c_s) \; . 
\ee
Using $c^2_s = dP/d \rho$, we can rewrite $c_s'$ as
\be \label{cs prime to g3}
c_s' = \frac{(1-g_3)}{2 c_s}  \; ,
\ee
so that the amplitude becomes 
\be \label{1to2 pp}
i {\cal M} = \frac{c_s p k}{\sqrt \rho}   \big[ \cos\theta +\sfrac12 (1-g_3) \big]  \; .
\ee
We recognize the same angular dependence as in \eqref{1to2 P(X)}, but with different overall normalization and dependence on $p$ and $k$. This had to be expected, for the two computations correspond to formally different processes from an $S$-matrix viewpoint: \eqref{1to2 P(X)} corresponds to a $1\to2$ process, whereas  \eqref{1to2 pp} corresponds to a $0\to1$ process.

To compare the two results we should compute the rate, which is physical and independent of how we normalize the states. Notice that in the point-particle theory, the only conservation delta function we have is
\be \la{cherenkov delta function}
(2\pi) \delta (\omega - \vec v_0 \cdot \vec k) \; .
\ee
This is because in the background of a particle moving at constant velocity, $\vec x_0 (t) = \vec v_0 t$, spatial and time-translations are broken down to this particular combination. This is made explicit by considering the structure of a generic term in the world-line action involving external fields as well as point particle degrees of freedom:
\begin{align}
\int dt f(\vec x_0(t), t) & = \int dt  \frac{d^3 k}{(2\pi)^3} \frac{d\omega}{(2\pi)} \, \tilde f(\vec k, \omega) \, e^{i \vec k\cdot \vec v_0 t }e^{-i \omega t}  \\
& = \int \frac{d^3 k}{(2\pi)^3} \frac{d\omega}{(2\pi)}
 \tilde f(\vec k, \omega) \, (2\pi) \delta(\omega - \vec v_0 \cdot \vec k) \; .
\end{align}
The   rate associated with \eqref{1to2 pp} thus is
\be
d \Gamma = |{\cal M}|^2 d\Pi_f =  \frac{c_s^2 p^2 k^2}{\rho}   \big[ \cos\theta +\sfrac12 (1-g_3) \big]^2 \times \frac{d^3 k}{(2\pi)^3 \, 2c_s k} \times (2\pi) \delta(\omega - \vec v_0 \cdot \vec k) \; ,
\ee
in perfect agreement with \eqref{emission}.

Consider now a slightly more complicated process:  elastic hard phonon-soft phonon scattering. 
To simplify the algebra somewhat, let's assume that we have an head-on collision\footnote{In systems with boost invariance, this corresponds to a choice of reference frame. Here however boost invariance is broken by the medium, and so this assumption corresponds to a specific choice of initial state.}, i.e. $\hat k = - \hat p$. 
 
For the $P(X)$ theory, the relevant diagrams are those of Fig.~\ref{scattering P(X)}, and we need the expansion of the action up to quartic order. In addition to the terms in \eqref{P(X) expanded}, we have
\be
\int d^4 x \, \frac{\rho}{c_s^2} \bigg\{ \frac{g_4}{4! c_s^4} \, \dot \pi^4 -\frac{1}{4 c_s^2}\, \dot \pi^2 (\vec \nabla \pi)^2 + \frac{1}{4!} \, (\vec \nabla \pi)^4 \bigg\} \; .
\ee
The tree-level scattering amplitude is
\be \label{amplitude scattering P(X)}
i {\cal M} = \frac{2 i c_s^2}{\rho} \times \frac{1-g_3 - g_3^2 + g_4-(1+g_3) \cos \theta}{1- \cos \theta} \times p^2 k^2 \; ,
\ee
where $k$ is the incoming soft momentum, $p$ is the incoming hard momentum, and $\theta$ is the angle between the momentum of the hard phonon and that of the outgoing soft one. We have also used that, by conservation of energy and momentum, the outgoing soft momentum is equal to
\be
k' \simeq \frac{2k }{(1-\cos\theta)} \; 
\ee
in the limit of small $k$ and $k'$ (compared to $p$).

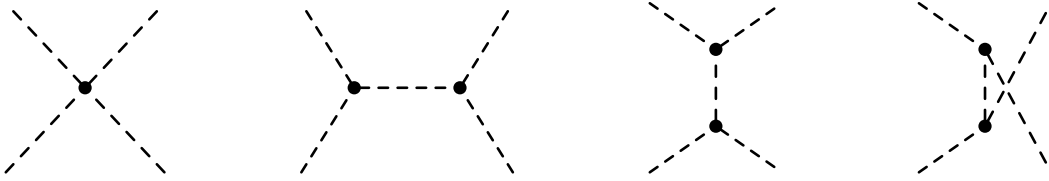
\begin{figure}[t]
\begin{center}
\subfigure
{
\begin{fmfgraph*}(60,80) 
\fmftop{o1,o2}
\fmfbottom{i1,i2}
\fmfdot{v}
\fmf{dashes}{i1,v,o2}
\fmf{dashes}{i2,v,o1}
\end{fmfgraph*}
}
\hspace{0.5in} 
\subfigure
{
\begin{fmfgraph*}(80,80) 
\fmftop{o1,o2}
\fmfbottom{i1,i2}
\fmfdot{v1}
\fmfdot{v2}
\fmf{dashes}{i1,v1,o1}
\fmf{dashes}{i2,v2,o2}
\fmf{dashes}{v1,v2}
\end{fmfgraph*}
}
\hspace{0.5in} 
\subfigure
{
\begin{fmfgraph*}(50,80) 
\fmftop{o1,o2}
\fmfbottom{i1,i2}
\fmfdot{v1}
\fmfdot{v2}
\fmf{dashes}{i1,v1,i2}
\fmf{dashes}{o1,v2,o2}
\fmf{dashes,tension=1.2}{v1,v2}
\end{fmfgraph*}
}
\hspace{0.5in} 
\subfigure
{
\begin{fmfgraph*}(50,80) 
\fmftop{o1,o2}
\fmfbottom{i1,i2}
\fmfdot{v1}
\fmfdot{v2}
\fmf{dashes}{i1,v1}
\fmf{phantom}{v1,i2}
\fmf{dashes}{o1,v2}
\fmf{phantom}{v2,o2}
\fmf{dashes,tension=1.2}{v1,v2}
\fmffreeze
\fmf{dashes}{v1,o2}
\fmf{dashes}{v2,i2}
\end{fmfgraph*}
}
\end{center}
\vspace{-0.5cm}
\caption{\small \it Lowest order contributions to phonon-phonon scattering in the $P(X)$ theory.} \la{scattering P(X)}
\end{figure}

On the other hand, for the point-particle theory, if we treat again the hard phonon as an external source, the diagrams are those in Fig.~\ref{scattering p.p.}, with the understanding that there is no propagator associated with the straight double lines. The third diagram arises because of a technical subtlety: even in the  $k \ll p$ limit, it is incorrect to treat the hard phonon as an external source as far as {\em intermediate} states are concerned. This is because in the point-particle theory expanded about the unperturbed trajectory $\vec x(t) = \vec v_0\,t$, the perturbations $\delta \vec x(t)$ of the trajectory are gapless---they are the Goldstone modes of spontaneously broken translations---and can thus be excited at arbitrarily low energies.
It turns out that their contribution as intermediate states to the amplitude under study is of the same order as the other contributions of fig.~\ref{scattering p.p.}, and should thus be kept (the same is true for, e.g., low frequency Compton scattering in QED.) The same holds for the fluctuations of $p$, since $p$ is one of the canonical conjugate variables of $\vec x$. That is why we also need to consider the third diagram in Fig.~\ref{scattering p.p.}, in which the wiggly double line represents the propagator of the fluctuations in the trajectory of the hard phonon. Notice finally that there is no point-particle analog of the fourth diagram in Fig.~\ref{scattering P(X)}. This is because in our point-particle approach the external hard phonon lines do not actually correspond to asymptotic states---they are just a visual aid to make the physical meaning of this diagram more transparent. 

In order to calculate the first two diagrams, we need to expand the point-particle action \eqref{Sphonon} up to quadratic order in $\pi$:
\be
S_{\rm phonon} \simeq \int dt \, p \l\{ v-c_s - c_s' \dot \pi + \hat v \cdot \nabla \pi + \sfrac{1}{2} \l[ \sfrac{1}{v} P_\perp^{ij} \d_i \pi \d_j \pi + c_s' (\nabla \pi)^2 - c_s'' \dot \pi^2\r] \r\},
\ee
where we have simplified the notation by defining $\vec v \equiv \dot{\vec x}$ and $ P_\perp^{ij} = \delta^{ij} - \hat v^i \hat v^j$. Setting $\vec v = c_s \hat v_0$, we find the following on-shell results for the first two diagrams in Fig.~\ref{scattering P(X)}:
\begin{subequations} \la{first two diagrams phonons}
\begin{align}
&\hspace{-0.8cm} \parbox{20mm}{
\begin{fmfgraph*}(50,40) 
\fmftop{i}
\fmfbottom{o}
\fmfdot{v}
\fmf{dbl_plain}{i,v,o}
\fmffreeze
\fmfright{o1,o2}
\fmf{dashes}{o1,v,o2}
\end{fmfgraph*}
}
= - \fr{2 i c_s p_0 k^2}{\rho} \l[ \frac{c_s' c_s \cos \theta + c_s'' c_s^3}{1-\cos \theta} \r]\\
&\hspace{-0.8cm} \!\!\!\!\!\! \parbox{20mm}{
\begin{fmfgraph*}(75,40) 
\fmftop{i}
\fmfbottom{o}
\fmfdot{v1}
\fmf{dbl_plain}{i,v1,o}
\fmffreeze
\fmfright{o1,o2}
\fmfdot{v2}
\fmf{dashes}{o1,v2,o2}
\fmf{dashes,tension=2.1}{v1,v2}
\end{fmfgraph*}
} 
\quad\, = \frac{i c_s p_0 k^2}{\rho} (1+c_s c_s') (c_s' c_s - \cos \theta) \frac{1 + \cos \theta}{(1-\cos \theta)^2}
\end{align}
\end{subequations}
Remember again that we are considering a head-on collision. It is worth to point out that the second diagram reproduces, up to an overall constant, the contribution coming from the second diagram in Fig. \ref{scattering P(X)}. This is known as the ``$u$-channel'' contribution in high energy physics parlance. This result should not be surprising. In fact, each of these two diagrams is given by the product of three factors---two vertices and one propagator. The propagators and the bulk vertices are identical for the two diagrams, and our discussion of phonon decay has shown that the world-line vertex reproduces the physics of the bulk vertex. The fact that the soft phonon interacting with the world-line is on-shell for the decay process and off-shell in the case of scattering is inconsequential: matching in effective field theories can always be performed  on- or off-shell~\cite{Rothstein:2003mp}.
\begin{figure}[t]
\begin{center}
\hspace{-1in} 
\subfigure
{
\begin{fmfgraph*}(100,80) 
\fmftop{i}
\fmfbottom{o}
\fmfdot{v}
\fmf{dbl_plain}{i,v,o}
\fmffreeze
\fmfright{o1,o2}
\fmf{dashes}{o1,v,o2}
\end{fmfgraph*}
}
\hspace{-1.5cm}
\subfigure
{
\begin{fmfgraph*}(150,80) 
\fmftop{i}
\fmfbottom{o}
\fmfdot{v1}
\fmf{dbl_plain}{i,v1,o}
\fmffreeze
\fmfright{o1,o2}
\fmfdot{v2}
\fmf{dashes}{o1,v2,o2}
\fmf{dashes,tension=2.1}{v1,v2}
\end{fmfgraph*}
}
\hspace{-.3cm}
\subfigure
{
\begin{fmfgraph*}(80,80) 
\fmftop{i}
\fmfbottom{o}
\fmfdot{v1}
\fmfdot{v2}
\fmf{dbl_plain}{i,v1}
\fmf{dbl_wiggly}{v1,v2}
\fmf{dbl_plain}{v2,o}
\fmffreeze
\fmfright{o1,o2}
\fmf{dashes}{o1,v2}
\fmf{dashes}{o2,v1}
\end{fmfgraph*}
}
\end{center}
\vspace{-0.5cm}
\caption{\small \it Lowest order contributions to phonon-phonon or roton-phonon scattering in the effective point-particle theory.} \la{scattering p.p.}
\end{figure}
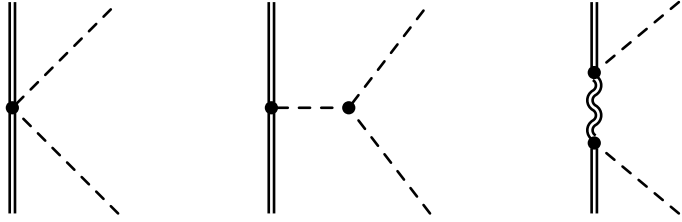

To compute the $\delta \vec x$ and $\delta p$ propagators that enter the third diagram in Fig. \ref{scattering p.p.}, we need the point-particle action expanded to quadratic order in $\delta \vec x$, $\delta p$ and to zeroth order in $\pi$:
\be
S_{\rm phonon} \supset \int dt \, \frac12 \Big[\frac{p_0}{c_s} P^\perp_{ij} \, \delta \dot {x}^i \delta \dot x^j + 2  (\hat v_0 \cdot \delta \dot {\vec x}) \, \delta p   \Big] \; .
\ee
The associated  propagators are
\be
\langle \delta p \, \delta p \rangle = 0 \; , \qquad \langle \delta p \, \delta \vec x \rangle =- \langle  \delta \vec x \, \delta p \rangle = -\frac{\hat v_0}{\omega}   \; , \qquad
 \langle \delta x^i \, \delta  x^j \rangle = \frac{c_s}{p_0} \frac{i}{ \omega^2} P_{ij}^\perp \; .
\ee
Moreover, to compute the new interaction vertices, we also need to keep all the $\delta \vec x$-$\pi$ bilinear terms:
\be
\int dt \l\{ \delta p (\hat v_0 \cdot \nabla \pi - c_s' \dot \pi ) + \frac{p_0}{c_s} P_\perp^{ij} \delta \dot x_i \d_j \pi - p_0 c_s' \delta \vec x \cdot \nabla \dot \pi + p_0 \hat v_0^i \delta x^j \d_i \d_j \pi \r\}. 
\ee
The new contribution to the amplitude thus is
\be
\parbox{20mm}{
\begin{fmfgraph*}(40,40) 
\fmftop{i}
\fmfbottom{o}
\fmfdot{v1}
\fmfdot{v2}
\fmf{dbl_plain}{i,v1}
\fmf{dbl_wiggly}{v1,v2}
\fmf{dbl_plain}{v2,o}
\fmffreeze
\fmfright{o1,o2}
\fmf{dashes}{o1,v2}
\fmf{dashes}{o2,v1}
\end{fmfgraph*}
}
\!\!\!\! = \frac{i c_s p_0 k^2}{\rho} (1 -c_s c_s') ( c_s' c_s + \cos \theta) \frac{1- 3 \cos \theta}{(1 -\cos \theta)^2}
\ee
Adding this result together with the diagrams in equations \eqref{first two diagrams phonons} we find the following total amplitude:
\begin{equation}
i \mathcal M = \frac{2 i c_s p_0 k^2}{\rho}  \frac{(c_s c_s' -1)\cos \theta + (c_s c_s')^2 -c_s^3 c_s''}{1 - \cos \theta}
\end{equation}
Using now the fact that
\begin{equation}
	c_s'' = - \frac{1}{4 c_s^3} \l( 1 - 3g_3^2 + 2 g_4\r)
\end{equation}
together with eq. \eqref{cs prime to g3}, we can rewrite this amplitude as
\begin{equation}
	i \mathcal M = \frac{i c_s p_0 k^2}{\rho} \times \frac{1-g_3 - g_3^2 + g_4-(1+g_3) \cos \theta}{1 - \cos \theta} .
\end{equation}
As we can see, this agrees with the result \eqref{amplitude scattering P(X)} we obtained from the $P(X)$ theory up to an overall normalization, as in the case of Cherenkov radiation. Once again, though, this discrepancy disappears once we calculate a physical quantity such as the cross section. The two theories require different kinematical factors to relate the amplitude  to the cross section, because the asymptotic states are different. Once such factors are taken into account, the cross sections are the same and are given by\footnote{\label{footnote: cross sections}
We are implicitly using the so-called relativistic normalization for asymptotic states (see related comments in \cite{...}). So, for the point-particle theory (a $1 \to 1$ process), we have 
$$ d \sigma = \fr{1}{2 \omega} \times \fr{1}{2 c_s} \times  |\mathcal{M}|^2 \times 2 \pi \delta (\omega_i - \omega_f - \vec v_0\cdot (\vec k_i - \vec k_f)) \times \fr{d^3 k_f}{(2 \pi)^3 2 \omega_f} \; , $$ 
where the conservation $\delta$-function is the appropriate one for world-line processes (see eq.~\eqref{cherenkov delta function}), whereas for the $P(X)$ theory (a $2 \to 2$ process) we have
$$
d \sigma = \fr{1}{(2 \omega)(2 c_s p)} \times \fr{1}{2 c_s} \times  |\mathcal{M}|^2 \times (2 \pi)^4 \delta (c_s p_i +\omega_i - c_s p_f - \omega_f) \delta^3 (\vec p + \vec k - \vec p_f - \vec k_f ) \times \fr{d^3 k_f}{(2 \pi)^3 2 \omega_f} \fr{d^3 p_f}{(2 \pi)^3 2 c_s p_f} \; .
$$}
\be
\frac{d \sigma}{d\Omega} = \frac{p^2 k^4}{32 \pi^2 c_s^2 \rho^2} \left[ \frac{1-g_3 - g_3^2 + g_4-(1+g_3) \cos \theta}{1 - \cos \theta} \right]^2 ,
\ee
with $p$ and $k$ denoting the incoming hard and soft momenta respectively. 

We thus see that, for kinematical configurations such that both effective theories can be applied, our effective point-particle  theory is equivalent to the more standard $P(X)$ effective field theory. The only technical subtlety one should consider is that fluctuations of the point particle's trajectory contribute as intermediate states in scattering amplitudes, even in the limit in which the point particle is much harder that the bulk modes it interacts with. We are now ready to apply this know-how to rotons, regarding which the $P(X)$ effective theory has nothing to say. We also refer the reader to Appendix \ref{generic}, where we sketch how to describe in our formalism the emission and scattering  of soft phonons by a generic particle or point-like object.

\section{Roton-phonon scattering} \label{sec:roton-phonon scattering}

Let us now turn our attention to the scattering of a soft phonon off a roton. In order to discuss this process, we will use the action \eqref{effective action for rotons2} where $\hat p$ has not been integrated out. The relevant diagrams describing this scattering process at lowest order are again the ones in Fig. \ref{scattering p.p.}, where now the double line stands for a roton rather than a hard phonon. In order to calculate these diagrams, we will need linear and quadratic couplings of the phonon field to the roton worldline. These can be easily obtained by expanding the Lagrangian \eqref{effective action for rotons2} in powers of $\pi$:
\begin{equation}
	L_{\rm roton} \simeq L_0 + L_1 + L_2 \; ,
\end{equation}
with
\begin{subequations}
\begin{align}
	L_0 &= - \Delta + p_* (\dot{\vec x} \cdot \hat p) + \frac{m_*}{2} (\dot{\vec x}\cdot \hat p)^2 \\
	L_1 &= \dot \pi L_0' + \vec \nabla \pi \cdot \vec p  \\
	L_2 &= \sfrac12  \dot \pi^2 L_0'' - \sfrac12 (\vec \nabla\pi)^2 L_0'  + \dot \pi \, \vec \nabla \pi \cdot \vec p^{\, \prime}  + \sfrac12 {m_*} (\vec \nabla \pi \cdot \hat p)^2,
\end{align}
\end{subequations}
We have simplified the notation by introducing the total momentum of the roton, $\vec p = (p_* + m_* \dot{\vec x} \cdot \hat p) \hat p$ and, as in the previous sections, primes denote derivatives with respect to $X$, evaluated on the $X= \mu/m$ background.

For a roton with
\be \label{roton solution}
\dot{\vec x}  = v_0 \,    \hat p_0 \; , \qquad \qquad \vec p = p_0 \, \hat p_0,
\ee 
we have
\begin{subequations}
\begin{align}
& \parbox{20mm}{
\begin{fmfgraph*}(50,40) 
\fmftop{i}
\fmfbottom{o}
\fmfdot{v}
\fmf{dbl_plain}{i,v,o}
\fmffreeze
\fmfright{or}
\fmf{dashes}{v,or}
\end{fmfgraph*}
}
= - \vec k \cdot \vec p_0 + \omega \bar L_0'\\
& \!\! \parbox{20mm}{
\begin{fmfgraph*}(58,40) 
\fmftop{i}
\fmfbottom{o}
\fmfdot{v}
\fmf{dbl_plain}{i,v,o}
\fmffreeze
\fmfright{o1,o2}
\fmf{dashes}{o1,v,o2}
\end{fmfgraph*}
}
 =  i \l[ \vec k_1 \cdot \vec k_2 \, \bar L_0' - \omega_1 \omega_2 \bar L_0'' - m_* (\vec k_1 \cdot \hat p_0)(\vec k_2 \cdot \hat p_0) +(\omega_1 \vec k_2 \cdot \vec p_0^{\, \prime} + \omega_2 \vec k_1 \cdot \vec p_0^{\, \prime}) \r] \; ,
\end{align}
\end{subequations}
where by convention all the $\omega$'s and $k$'s are incoming (and related by the conservation delta-function \eqref{cherenkov delta function}), 
and we have denoted by $\bar L_0$ the on-shell value of $L_0$ calculated on the solution \eqref{roton solution}. 

As remarked in the previous section, in general we also need to consider contributions coming from intermediate fluctuations of the point-particle trajectory. For the scattering process we are considering, these effects are captured by the last diagram in Fig. \ref{scattering p.p.}. 
In order to calculate it, we need the propagator for the fluctuations $\delta \vec x$ and $\delta \hat p$ of the trajectory. To this end, we expand $L_0$ up to quadratic order to find
\begin{equation} \label{Quadratic Lagrangian fluctuations rotons}
	L_0^{(2)} \simeq \sfrac12 {m_*} (\delta \dot{\vec x}\cdot \hat p_0)^2 + p_0 \, \delta \dot{\vec x}\cdot \delta \hat p - \sfrac12 {p_0 v_0} \, \delta \hat p \cdot \delta \hat p.
\end{equation}
We then invert this kinetic term to find the propagator, keeping in mind that $\delta \hat p$ contains only two independent components 
because $\hat p$ has unit norm. It is easier to work with a slightly redundant parametrization of the propagator that consists of a $6\times 6$ matrix acting on the $( \delta \vec x, \delta \hat p)$ space and satisfying the constraints
\begin{equation}
\hat p_0^i \langle \delta \hat p_i \delta \hat p_j \rangle =0, \qquad \qquad \hat p_0^i \, \langle \delta \hat p_i \delta  x_j \rangle = \hat p_0^i \, \langle \delta  x_j  \delta \hat p_i \rangle =0 \; ,
\end{equation}
which, to the order we are working, implement the constancy of $\hat p \cdot \hat p$.
Then, the propagator for the fluctuations of the roton trajectory is
\begin{equation}
\parbox{20mm}{
 \begin{fmfgraph*}(50,20) 
\fmfleft{i}
\fmfdot{i}
\fmfright{o}
\fmfdot{o}
\fmf{dbl_wiggly}{i,o}
\end{fmfgraph*}
}
= \l(\begin{tabular}{cc}
 $ \frac{i v_0 P^{ij}}{\omega^2 p_0} + \frac{i \hat p_0^i \hat p_0^j}{\omega^2 m_*}$ & $- \frac{P^{ij}}{\omega p_0}$\\
 $ \frac{P^{ij}}{\omega p_0} $ &  0
\end{tabular} \r), \qquad\qquad  P^{ij} = \delta^{ij} - \hat p_0^i \, \hat p_0^j.
\end{equation}
The Feynman rules for the  vertices that appear in the third diagram are obtained by expanding $L_1$ up to linear order in $\delta \vec x $ and $\delta \hat p$. In Fourier space, this yields
\begin{equation}
 \parbox{20mm}{
\begin{fmfgraph*}(50,40) 
\fmftop{o}
\fmfbottom{i}
\fmfdot{v}
\fmf{dbl_plain}{i,v}
\fmf{dbl_wiggly}{v,o}
\fmffreeze
\fmfright{or}
\fmf{dashes}{v,or}
\end{fmfgraph*}
}
 =  \l( \!\! \begin{tabular}{c} $- i \vec k  ( \vec k\cdot \vec p_0 - \omega \bar L_0' ) + i \tilde \omega \hat p_0 ( m_* \vec k \cdot \hat p_0 - \omega  p_0' ) $ \\ $ - p_0 \vec k$\end{tabular} \! \r),
\end{equation}
where  $\tilde \omega \equiv \omega - \vec v_0 \cdot \vec k$ is the frequency of the $\delta \vec x$, $\delta \hat p$ perturbation, and the phonon's frequency and momentum are again incoming.
Physically, this diagram describes the perturbation that an incoming phonon induces on the trajectory of a roton.

We are now ready to combine all the ingredients we have derived so far and use them to calculate the Feynman diagrams in Fig. \ref{scattering p.p.}. Denoting with $\omega_i$ and $\vec k_i$ ($\omega_f$ and $\vec k_f$) the frequency and momentum of the incoming (outgoing) phonon, and with $\vec q \equiv \vec k_i - \vec k_f$ the momentum transfer, we find 
\begin{subequations}
\begin{align}
&\hspace{-0.8cm} \parbox{20mm}{
\begin{fmfgraph*}(50,40) 
\fmftop{i}
\fmfbottom{o}
\fmfdot{v}
\fmf{dbl_plain}{i,v,o}
\fmffreeze
\fmfright{o1,o2}
\fmf{dashes}{o1,v,o2}
\end{fmfgraph*}
}
= \fr{i c_s^2}{\rho} \l[\omega_f \omega_i \bar L_0'' - \vec k_f \cdot \vec k_i  L_0' +m_* (\vec k_i \cdot \hat p_0) (\vec k_f \cdot \hat p_0) - (\omega_i \vec k_f \cdot \vec p_0^{\, \prime} +\omega_f \vec k_i \cdot \vec p_0^{\, \prime}  ) \r] \\
&\hspace{-0.8cm} \!\!\!\!\!\! \parbox{20mm}{
\begin{fmfgraph*}(75,40) 
\fmftop{i}
\fmfbottom{o}
\fmfdot{v1}
\fmf{dbl_plain}{i,v1,o}
\fmffreeze
\fmfright{o1,o2}
\fmfdot{v2}
\fmf{dashes}{o1,v2,o2}
\fmf{dashes,tension=2.1}{v1,v2}
\end{fmfgraph*}
} 
\quad\, = -\fr{2i}{\rho} \l(\vec k_i \cdot \vec k_f + \fr{c_s'}{c_s} \omega_i \omega_f \r)(\vec v_0 \cdot \hat q )(\bar L_0' \vec v_0 - \vec p_0) \cdot \hat q \la{roton diagram 2}\\
 &\hspace{-0.8cm} \,\,\,\parbox{20mm}{
\begin{fmfgraph*}(40,40) 
\fmftop{i}
\fmfbottom{o}
\fmfdot{v1}
\fmfdot{v2}
\fmf{dbl_plain}{i,v1}
\fmf{dbl_wiggly}{v1,v2}
\fmf{dbl_plain}{v2,o}
\fmffreeze
\fmfright{o1,o2}
\fmf{dashes}{o1,v2}
\fmf{dashes}{o2,v1}
\end{fmfgraph*}
}
\!\!= - \fr{i c_s^2}{\rho} \bigg\{ \vec k_i^\perp \cdot \vec k_f^\perp \l[ \fr{v_0}{\omega^2 p_0} (\vec k_f \cdot \vec p_0 - \omega_f \bar L_0')(\vec k_i \cdot \vec p_0 - \omega_i \bar L_0') + \fr{(\vec k_i + \vec k_f) \cdot (\vec p_0 - \vec v_0 \bar L_0')}{\omega}  \r] \nonumber \\
& \qquad \qquad \qquad + \fr{1}{\omega^2 m_*} \l[(\vec k_f \cdot \vec p_0 - \omega_f \bar L_0') (\vec k_f \cdot \hat p_0) + \tilde \omega (m_* \vec k_f \cdot \hat p_0 - \omega_f p_0')  \r] \\
& \qquad \qquad \qquad\qquad \qquad\qquad \times \l[(\vec k_i \cdot \vec p_0 - \omega_i \bar L_0') (\vec k_i \cdot \hat p_0) + \tilde \omega (m_* \vec k_i \cdot \hat p_0 - \omega_i p_0')  \r] \bigg\} \; , \nonumber 
\end{align}
\end{subequations}
where  $\vec k^\perp$ is the component of $\vec k$ perpendicular to $\vec v_0$. Notice also that, like before, there is a factor of $c_s/\sqrt{\rho}$ associated with each external phonon line.

We can now use the total amplitude---obtained by adding the three results above---to calculate the scattering cross section. For simplicity, we will restrict ourselves to a process in which the roton is at rest, $v_0 = 0$. In this case, the diagram in eq. \eqref{roton diagram 2} vanishes, and the phonon frequency is conserved:
\be
k_i = k_f \equiv k \; , \qquad \omega_i = \omega_f = c_s k \; .
\ee 
We should stress once more that this is not just an innocuous choice of reference frame: since boosts are spontaneously broken by the medium, the scattering amplitude depends on the velocities of the excitations with respect to the medium, and not just on their relative velocity like in a scattering process in vacuum. In order to make the comparison with earlier results~\cite{landau1949theory} easier, we will trade $X$-derivatives for derivatives with respect to the density,
\begin{equation}
	\frac{d}{d X} = \frac{d P}{d X} \frac{d \rho}{d P}  \frac{d}{d\rho} = \frac{\rho}{c_s^2} \frac{d }{d \rho}.
\end{equation}
Then, the total amplitude reduces to
\begin{align} \la{iM roton phonon}
i \mathcal M & \big|_{v_0 = 0} =  \nonumber \\
& - \frac{i c_s p_* k^2}{\rho} \bigg\{ (\hat k_i \cdot \hat k_f)	(\hat k_i + \hat k_f)\cdot \hat p_0 + \frac{p_*}{m_* c_s} (\hat k_i \cdot \hat p_0)^2 (\hat k_f \cdot \hat p_0)^2 + \frac{\rho^2}{c_s p_*}\bigg[ \frac{d^2 \Delta}{d \rho^2} +\frac{1}{m_*} \Big(\frac{d p_*}{d \rho} \Big)^2 \bigg]  \nonumber \\
& \qquad \qquad  \quad   +   B \cdot \frac{\rho}{p_* c_s}\frac{d \Delta}{d \rho}- \frac{\rho }{m_* c_s} \frac{d p_*}{d \rho} \l[ (\hat k_i \cdot \hat p_0)^2 + (\hat k_f \cdot \hat p_0)^2  \r] \bigg\}, 
\end{align}
where
\begin{subequations}
\begin{align}
	B &\equiv 1 - 2 \frac{\rho}{c_s} \frac{d c_s}{d \rho} + (\hat k_i \cdot \hat p_0)(\hat k_f\cdot \hat p_0) \bigg[ 2 + \frac{\rho}{m_* c_s^2}\frac{d \Delta}{d \rho} +\frac{p_*}{m_* c_s} (\hat k_i + \hat k_f)\cdot \hat p_0\bigg]   \\
	& \qquad \qquad \qquad \qquad \qquad \qquad \qquad \qquad \qquad \qquad \qquad \qquad -\frac{\rho}{m_* c_s}\frac{d p_*}{d \rho} (\hat k_i + \hat k_f)\cdot \hat p_0 \; . \nonumber
\end{align}
\end{subequations} 
The differential cross section is given in terms of this amplitude by\footnote{This cross-section is defined similarly to the one for phonon-phonon scattering (see footnote \ref{footnote: cross sections}), except that here the relative velocity in the initial state is $c_s$ rather than $2 c_s$.}
\begin{align}
\fr{d\sigma}{d\Omega} = \fr{| \mathcal{M}|^2}{16 \pi^2 c_s^4}.
\end{align}  

The cross section for roton-phonon scattering was first calculated by Landau and Khalatnikov (LK)~\cite{landau1949theory} in the limit of small $d \Delta / d \rho$. The first line in our total amplitude \eqref{iM roton phonon} reproduces exactly their result. However, the second line of our amplitude introduces some corrections: while the first term, proportional to $d \Delta / d \rho$, was consistently neglected in LK's approach, the second term should have been kept. We are quite confident of this result, since we have derived it also starting from the action in eq. \eqref{effective action for rotons1}, as opposed to the one in eq. \eqref{effective action for rotons2} used in this section. The action \eqref{effective action for rotons1} leads to different Feynman rules, but the final result for the total amplitude remains the same. Using standard kinetic theory arguments, one can relate the phonon-phonon and roton-phonon cross sections to macroscopic observables, such as the temperature dependence of viscosity~\cite{landau1949theory2}. Hence, our corrections have potentially observable consequences.

\section{On floating and sinking}\label{gravity}

Let us now turn our attention to a question that is probably quite natural for a reader with a high energy background, but it's admittedly a bit unusual in a condensed matter context: how does gravity act on a medium's excitations, such as phonons and rotons? For ordinary objects immersed in a generic fluid we have the Archimedean principle, but it is not obvious how to apply that to more general ``objects": what is the volume displaced by a phonon or a roton? And what are their gravitational masses?

The effect of gravity on sound waves can easily be understood using heuristic arguments based on classical wave mechanics.\footnote{We thank Eric Cornell for discussions about this point.} An external gravitational field gives rise to a pressure gradient in the fluid, which in turn induces a gradient in the sound speed. Consider now a wavepacket propagating along a surface of constant pressure. The upper and lower parts of this wavepacket will move at slightly different speeds, and as a result its trajectory will bend in the direction opposite to that of the sound speed gradient.
The sound speed is usually larger in regions of larger pressure\footnote{For instance, for zero-temperature liquid helium $c_s \sim 1/ m a \propto \rho^{1/3}$, with $a$ the interparticle separation, and thus 
\begin{equation*}
	\frac{d c_s^2}{d P} = \frac{1}{c_s^2} \frac{d c_s^2}{d \rho} > 0.
\end{equation*}}, 
and so sound will tend to float rather than sink. 
It is thus natural to expect that phonons---the quanta of sound---want to float.
But what about inherently quantum mechanical excitations such as rotons? 

The effective point-particle theory  developed in Sec.~\ref{effective} is perfectly suited to tackle this question. Essentially, this is because it is constructed starting from considerations involving spacetime symmetries, and can thus be extended straightforwardly to incorporate gravitational phenomena, since gravity is the gauge field for spacetime symmetries. As shown in detail in Appendices \ref{appb} and \ref{appc}, the effect of an external gravitational potential $\Phi(x)$ on a non-relativistic superfluid and on the particles living in it is captured by a simple $\Phi$-dependent shift of the chemical potential, or, equivalently, of our variable $X$:
\begin{equation} \label{gravitational coupling}
	X = {\mu}/{m} + \dot \pi - \tfrac{1}{2} (\vec \nabla \pi)^2 - \Phi.
\end{equation}
For a reader familiar with trapped superfluids, this shift of the chemical potential will be reminiscent of the way in which trapping potentials are usually included in the Gross-Pitaevskii model (see e.g. \S 4.4 of~\cite{Schakel:2008zz}).\footnote{We are grateful to Angelo Esposito for pointing this out to us.} Indeed, a gravitational potential can always be interpreted as a (admittedly, quite weak) trapping potential of sorts. However, we should stress that the rule-of-thumb for trapping potentials is derived at weak coupling---which is the regime in which Gross-Pitaevskii is applicable---whereas \eqref{gravitational coupling} is completely general and can be derived without making any assumption about the underlying physics that gives rise to the superfluid state.

For what follows, it is also instructive to realize that, when applied to a superfluid at rest, the prescription \eqref{gravitational coupling} is nothing but a rewriting of hydrostatic equilibrium in a static gravitational field. In fact, setting the perturbations of the superfluid to zero, $\pi =0$, we have $X = \mu/m -\Phi(\vec x)$. According to our effective theory the superfluid density and pressure are $\rho = P'(X)$  and $P=P(X)$ (see eq. (\ref{p rho u})), and thus it follows immediately that
\begin{equation}
	\vec \nabla P =  - \rho \vec \nabla \Phi \; ,
\end{equation}
which is the hydrostatic equilibrium condition. This means that, when applied to the effective point-particle theory for particles living in the superfluid, the shift \eqref{gravitational coupling}  describes the {\em net} effect of gravity onto these particles: it includes the direct gravitational pull as well as the buoyant force of Archimedean fame, because the pressure gradients induced by gravity are automatically taken into account.

Let's first see explicitly how this works in the case of an ordinary object,  which we are all familiar with. Performing the replacement \eqref{gravitational coupling} in the action \eqref{Sobj}, setting to zero the $\pi$ perturbations, and expanding to first order in $\Phi$ we get
\be
S_{\rm obj} \to \int dt \big[E_0' \, \Phi + \sfrac12 M_{\rm eff} { \xdot \,^2} + \dots \big] \; ,
\ee
where the prime denotes a derivative w.r.t.~$X$ and we neglected terms of order $\Phi \dot x ^2$, which in the non-relativistic limit are subleading compared to those that we have kept. The equation of motion for our particle reads
\be
\frac{d \vec p}{d t} =  E_0' \vec \nabla \Phi \; ,   \qquad \vec p = M_{\rm eff} \xdot \; .
\ee
Comparing this to the Archimedean principle, we see that $E_0(X)$ must be such that
\be \label{E0'}
E_0'(X)= - \big(M(X) - V(X) \rho(X) \big) \; ,
\ee
or, equivalently,
\be
E_0(X)= - \int dX \, \big( M(X) - V(X) \rho(X) \big) \; ,
\ee
where $M$ is the mass of the object, and $V$ the  volume displaced, both evaluated at the local value of $X$ (we keep a potential $X$-dependence in $M$ for reasons that will be relevant in the example below.) However, unless we know in advance the mass and the displaced volume for our object, our effective point particle theory shows that in general there is no clear physical distinction between the two contributions to the force: the two quantities only appear in the combination $E'_0(X)$. This is also related to the manipulations that we performed in Eq.~\eqref{manipulate}: a mass parameter for the standard, empty space-like kinetic and gravitational energies, eqs.~\eqref{v^2} and \eqref{M Phi}, can be completely reabsorbed into our more general structure \eqref{Spp}.

Perhaps the following example will clarify the arbitrariness of the gravitational/buoyant  splitting of the net  force: consider a sponge. In empty space, when it's dry, it's easy to determine its mass, but essentially impossible to determine its ``solid" volume. Once we immerse it water and it gets completely soaked, the Archimedean principle formally applies, but in practice we do not know what the displaced volume is. There is a completely equivalent description in which we never talk about the dry sponge in empty space, but only about the soaked one in water: we can assign a net mass to it (sponge + soaked up water), which we do not know how to determine, a displaced volume, which now is easy to compute in principle---say the volume of a parallelepiped if the sponge has that macroscopic shape---and again the Archimedean principle applies. Clearly, an experimenter who never had access to the dry sponge and who does not have enough spatial resolution to determine that it's made of porous material, will adopt this second viewpoint. In going from one description to the other the mass of the water soaked up by the sponge moves around as far as the Archimedean principle is concerned: it moves from the buoyant term to the gravitational term. But the net buoyancy, which is the only measurable thing {\em in water}, remains the same.

As a further check, consider the gravitational field {\em produced} by an object immersed in a fluid. To simplify the discussion, consider the case in which we have a big self-gravitating sphere of an incompressible fluid of given density $\rho$, and we place at its center a much smaller rigid sphere of total mass $M$ and volume $V$. The Poisson equation in the presence of the object reads
\begin{align}
\nabla^2 \Phi & =  4\pi G \big( \rho \, \theta(r-R) + \rho_{\rm obj} \, \theta(R-r)  \big)\\
& =  4\pi G \big( \rho + (\rho_{\rm obj} - \rho) \, \theta(R-r)  \big)\; ,
\end{align}
where $\rho_{\rm obj}$ is the density of our object, and $R$ its radius.
In the point-particle limit, we can perform the replacement
\be
(\rho_{\rm obj} - \rho)  \theta(R-r) \to  \delta^3(\vec x) (M - \rho V) \; ,
\ee
as can be seen from integrating both sides in $d^3 x$. The Poisson equation thus reduces to
\be
\nabla^2 \Phi = 4\pi G \big( \rho +  (M - \rho V) \delta^3 (\vec x) \big) \; ,
\ee
from which we see that, even as far as the production of gravitational fields goes, the coupling of our material object to gravity is determined solely by the combination   \eqref{E0'}, with no physically relevant distinction between the two individual contributions.

All this is reassuring for us, because for more general ``objects" such as our excitations, there are no obvious candidates for masses and displaced volumes. Fortunately, all the information that we need is already contained in the actions that we wrote down. In particular, in the presence of gravity we will derive  for our excitations equations of motion of the form
\be
\frac{d \vec p}{dt} =  - m_g \vec \nabla \Phi \; ,
\ee
where $\vec p$ is the excitation's momentum, and $ m_g$ a quantity playing the role of what $-E_0'$ is for an ordinary object. This follows simply from the fact that, to first order in $\Phi$, our Hamiltonians have the general structure $H(\vec p, \vec x) = H_0(|\vec p \, |) + m_g (p)\Phi(\vec x)$. Given the discussion above, we will call $ m_g$ the ``net gravitational mass'' of the excitation in question.

\subsection{Phonons}
For phonons, performing the replacement \eqref{gravitational coupling} and setting to zero the $\pi$ perturbations, we get
\begin{equation}
	S_{\rm phonon} \to \int dt \, p  \big[ \, | \xdot| - c_s (\mu/m - \Phi) \big] \; .
\end{equation}
By varying this action w.r.t. $p$ we recover the constraint that the phonon must move at the speed of sound, which now depends on the position if $\Phi$ is not homogeneous. The variation w.r.t. $\vec x$ yields instead
\begin{equation} \label{sound bending}
	\fr{d \vec p}{dt} =  p \, c_s' \vec \nabla \Phi, \qquad \qquad \quad \vec p \equiv p \, \dot{\vec{x}} / |\dot{\vec{x}}| \; .
\end{equation} 
Thus, the net  gravitational mass of a phonon of momentum $p$ is $ m_{ g} = - p c_s'$, and its sign determines whether the phonon tends to sink (positive) or float (negative). Using $c^2_s = P' / P''$, and $\rho = P'$, we have
\begin{equation} \label{cs prime}
	c_s' = \frac{d\rho}{dX} \frac{dc_s}{d\rho} = \rho \,  \frac{d  c_s}{d  P} \; ,
\end{equation}
where the derivatives are understood to be taken at constant temperature (zero, in our case). Experimental data show that the sound speed of superfluid helium-4 at low temperatures increases with pressures~\cite{gibbs1999collective}. This implies that phonons have negative  net gravitational mass, and thus tend to float, in agreement with the  heuristic argument above. Notice that the effect is small, but not incredibly so: on dimensional grounds one has $c_s' \sim 1/c_s \sim 10^{-2}$ s/m, and so a phonon traveling horizontally in the Earth's gravitational field ($\nabla \Phi \sim 10$ m/s$^2$) tends to bend upwards at a rate
\be
\frac{d \theta}{dt} \sim (10 \; {\rm s})^{-1}\; ,
\ee
where $\theta$ is the angle with the horizontal.

It is interesting to compare our effect to the standard refraction of sound waves in inhomogeneous media. For so-called stratified media ($c_s=c_s(z)$, with $z$ being the vertical), refraction is usually phrased in terms of Snell's law (see e.g.~\cite{whitham2011linear}),
\be
\frac{\cos \theta}{c_s(z)} = {\rm const} \; ,
\ee 
where $\theta$ is again the angle with the horizontal. It is a matter of simple algebra to see that our eq.~\eqref{sound bending} implies such a conservation law. In fact, this is just a combination of the conservation laws for energy, $c_s(z) p = {\rm const}$, and for horizontal momentum, $p \cos \theta = {\rm const}$, which apply because time translations and horizontal spatial translations are unbroken.  We thus see that our effect is nothing but  ordinary refraction in disguise. We find it interesting that  within our formalism this phenomenon is gravitational in origin, in the sense that it encodes the net effect of gravity onto phonons, formally on the same footing as the net effect of gravity (gravitational force + buoyant force) onto ordinary objects. We are not sure whether this hints at something deep or trivial.

It is also interesting to compare our eq. \eqref{sound bending} to the relativistic phenomenon of light bending in vacuum. According to general relativity, the gravitational mass of a photon in the weak field limit is also proportional to its momentum, and its equation of motion reads~\cite{Ohanian:1995uu}
\begin{equation}
	\fr{d \vec p}{dt} \simeq - \frac{2 p}{c} \vec \nabla \Phi.
\end{equation}
Thus, the gravitational mass of photons is positive, which is why light bends towards rather than away from massive objects. In other words, while pho\emph{n}ons rise in a gravitational field, pho\emph{t}ons fall in accordance with the equivalence principle. The effect on photons is also much weaker, by a factor $c_s/c \sim 10^{-6}$. What a difference a single letter can make!


\subsection{Rotons}
Let us now turn our attention to rotons. A roton in a static superfluid placed in a gravitational field is described by the action \eqref{effective action for rotons1} with $\vec u = 0$ and $X = \mu/m- \Phi$, i.e.
\begin{equation}
   S_{\rm roton}^{R,L} \simeq \int dt \, \big[ -\Delta(\mu/m-\Phi) \pm p_*(\mu/m-\Phi) |\xdot| + \sfrac12 m_*(\mu/m-\Phi) |\xdot|^2 \big] \; .
\end{equation} 
Recall that the $+$ ($-$) sign in the Lagrangian refers to rotons with momenta on the right (left) of the roton minimum. Varying this action w.r.t. $\vec x$, we find
\begin{equation} \label{dpdt roton}
	\fr{d \vec p}{dt} = \Delta' \, \vec \nabla \Phi \, , \qquad \vec v \equiv \xdot \; , \quad \vec p \equiv (\pm p_* + m_* v) \hat v \; , 
\end{equation}
where we neglected terms of order $\Phi v$ and $\Phi v^2$, since for rotons $v \ll c_s$. The net gravitational mass of a roton thus is
$m_{g} = - \Delta'$, which following the same manipulations as above we can rewrite as
\begin{equation}
	m_g = - \rho \frac{d\Delta}{dP} \; .
\end{equation}
Experimental results for helium-4 show that this is positive~\cite{bartley1974pressure}, suggesting that rotons tend to sink. However, given the unconventional relationship between momentum and velocity, determining the actual trajectory of a sinking roton can be quite complicated. 

\begin{figure}[t]
\begin{center}
\includegraphics[width=14cm]{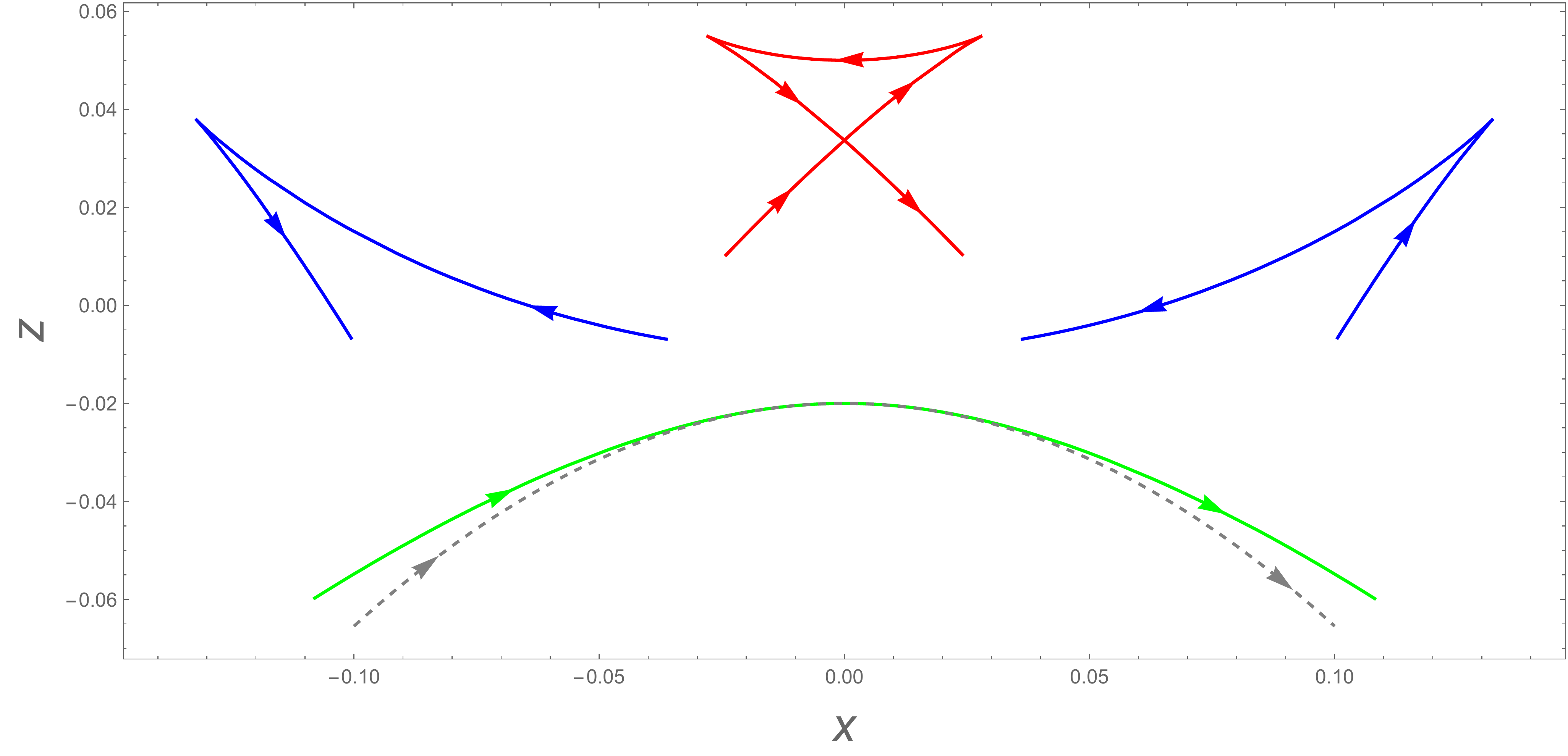}
\end{center}
\vspace{-0.5cm}
\caption{\small \it Possible trajectories for rotons in a vertical gravitational field $\vec g = -g \hat z$. For all the trajectories displayed, the horizontal component of the momentum, which is conserved, is directed towards the right. The arrows point in the direction of forward time evolution. We truncated the trajectories in the past and in the future when the rotons get too far from the roton minimum for our approximations to apply, which for definiteness we characterized by $|p-p_*|/p_*>30\%$. The coordinates are in units of $p_*^2/{m_* m_g g }$. According to the classification in the main text, we have green $=$ no turning points ($p_x = 1.1 \, p_*$); red $=$ two turning points ($p_x = 0.9 \, p_*$); blue, right $=$ one turning point $+$ metamorphosis ($p_x = 0.69 \, p_*$); blue, left $=$ vice-versa (ditto). For comparison, we are showing in dashed grey the trajectory of a normal object with parameters such that the curvature and velocity at the top are the same as for our no turning point roton trajectory.} \la{crazy}
\end{figure}

In practice, it is easier  to first solve for the momentum as a function of time, and then to integrate eq.~\eqref{velocity roton} to find the trajectory. For a constant gravitational acceleration $  \vec g \equiv -\vec \nabla \Phi$, the momentum is simply
\be \label{p(t)}
\vec p \,(t) = \vec p_0+m_g \vec g \, t \; ,
\ee
where $\vec p_0$ is the initial momentum. The velocity thus is
\be \label{xdot sink}
\xdot(t) = \frac{\vec p_0+m_g \vec g \, t}{m_*} \, \Big( 1 - \frac{p_*}{|\vec p_0+m_g \vec g \, t|}\Big) \; ,
\ee
and this can be integrated analytically to yield  $\vec x(t)$. The actual expression is not particularly illuminating\footnote{With $p_x$ in units of
$p_*$, $t$ in units of $m_g g / p_*$, and $x$ and $z$ in units of $m_* m_g g / p_*^2 $, the solution for motion in the $(x,z)$ plane reads:
\begin{equation}
	 x( t) =  p_x   t -  p _x \log\frac{t+\sqrt{ p_x^2 +  t^2}}{p_x} \; , \qquad   z( t) = - t^2/2 +(\sqrt{ p_x^2+ t^2}- p_x) \; .
\end{equation}
}, but for generic initial $\vec p_0$ the  trajectory can be quite spectacular, with the roton initially exhibiting an erratic behavior before deciding to sink to the bottom---see fig.~\ref{crazy}.

One can gain some intuition into such a peculiar behavior by noticing that, depending on the relative direction of $\vec p_0$ and $\vec g$, and on whether $p_0$ is bigger or smaller than $p_*$, the velocity \eqref{xdot sink} can end up crossing zero. This happens whenever $|\vec p_0+m_g \vec g \, t|$ crosses $p_*$. At that moment the direction of $\xdot$ relative to that of $\vec p$ changes sign, and, since this happens while $\vec p$ is nonzero and evolving smoothly with time, this corresponds to a turning point in the trajectory. In other words: $\vec p \, (t)$ is smoothly ``sinking" as in \eqref{p(t)}, but by doing so it can make the roton move between the left and the right of the roton minimum, thus inverting the sign of its velocity. 

\begin{figure}[t]
\begin{center}
\includegraphics[width=10cm]{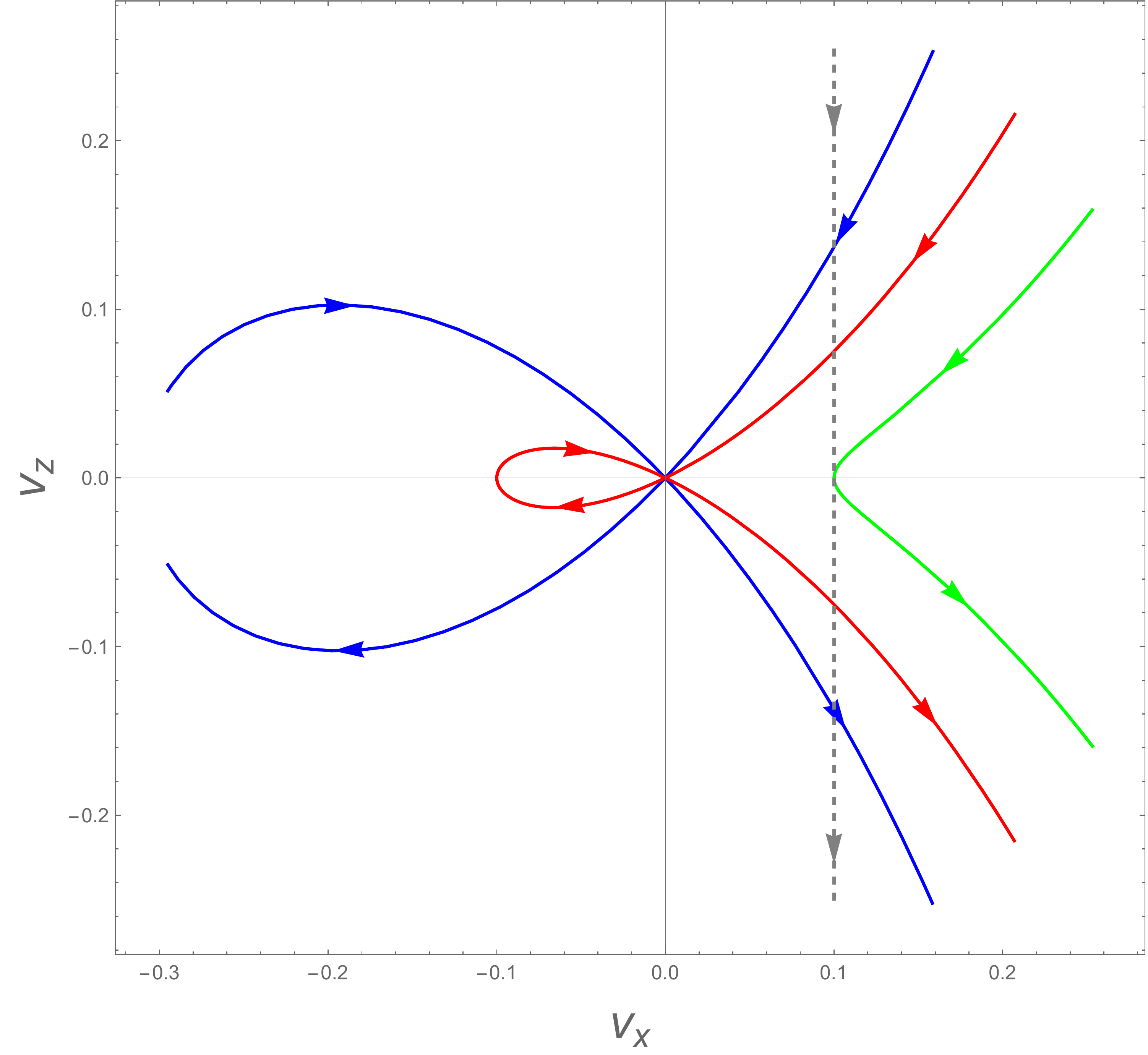}
\end{center}
\vspace{-0.5cm}
\caption{\small \it Possible velocity-space trajectories for rotons in a vertical gravitational field. The same conventions as in fig.~\ref{crazy} apply. The velocities are in units of $p_*/{m_* }$.} \la{velocity space}
\end{figure}

Specifically, imagine extrapolating the trajectories to all times, past and future. The time-evolution of the momentum, eq.~\eqref{p(t)}, happens on a plane, and so does that of the velocity, which is aligned or anti-aligned with the momentum. We can thus restrict to motion in the $x$-$z$ plane without loss of generality. Then, in going from $t= -\infty$ to $t=+\infty$, $p_z(t)$ spans all possible values, whereas $p_x$ is conserved, $p_x(t) = p_{0,x}$ for all $t$'s. We can thus classify these extrapolated orbits by their $p_x$, which without loss of generality we assume to be positive. Since $|\vec p \, | = \sqrt{p_x^2 + p_z^2}$ becomes large at large positive and negative times, because $p_z$ does, we see that all orbits start and end on the right of the roton minimum. We also have that $|\vec p \, | \ge p_x$ for all times. Keeping in mind that our approximations break down when we get too far from the roton minimum, we thus see that trajectories can feature: 
\begin{itemize}
\item No turning points: This happens for $p_x > p_*$. In this case $|\vec p \, (t)|$ is always on the right of the roton minimum.
\item Two turning points: This happens for $p_x < p_*$, but with $(p_* - p_x) \ll p_*$. In this case $|\vec p \, (t)|$ momentarily drops below $p_*$ at intermediate times, while staying always close to the roton minimum.
\item One turning point + Metamorphosis (or viceversa): This happens for $p_x < p_*$, but with $(p_* - p_x) \sim p_*$. In this case $|\vec p \, (t)|$ drops below $p_*$ at intermediate times, but it  decreases so much that it ends up violating the condition $p_* -|\vec p \, | \ll p_*$. At that point we are far from the roton minimum, and our approximations break down. The roton has effectively turned into something else. The reverse can also happen:  a roton is created at time $t=0$ with overall momentum $|\vec p \, | < p_*$, with negative $p_z$, and with $p_x$ substantially smaller that $p_*$; the extrapolation {\em back} in time of its trajectory at some point would take it too far to the left of the roton minimum for our approximations  to be valid; however, the forward time-evolution is within the regime of validity of our approximations, and makes the roton experience one turning point and then sink.
\end{itemize}
Of course our approximations also break down at  large times, both in the past and in the future, because the momentum becomes much bigger than $p_*$ then. So, overall, during the lifespan of our roton, we can have zero, one, or two turning points in its trajectory.
All these possibilities are depicted in position space and in velocity space  in figs.~\ref{crazy} and \ref{velocity space}.

On dimensional grounds, the timescale for all these phenomena to happen in the Earth's gravitational field is roughly the same as the one relevant for phonon bending,
\be
\tau \sim 10 \; {\rm s} \; .
\ee
This is because $m_g = -\Delta' \sim m_*$, and $\vec p$ changes by $p_*$ (in direction or magnitude) over a time $\tau \sim p_*/m_g g \sim c_s/g$.
However, for an initially very slow roton with momentum $p_0 = p_*(1+\epsilon)$ forming a  small (positive) angle $\theta_0 \gtrsim \sqrt{\epsilon}$ with the horizontal, the two-turning point phenomenon happens on a parametrically shorter timescale, $\sqrt{\epsilon} \, \tau$, because the momentum only has to change by $\delta \vec p \sim p_* \sqrt{\epsilon}$. The typical velocities during this period are of order 
$v \sim \epsilon c_s$, corresponding to typical displacements of order $\delta x \sim \epsilon^{3/2} c_s \tau \sim \epsilon^{3/2}$ km, unfortunately still too large to be relevant for experiments unless $\epsilon$ is extremely small.


\section{Discussion and outlook} \label{sec:conclusions}

We have introduced an effective point-particle formalism to describe how particles---which could be actual particles, collective excitations, or macroscopic objects alike---couple to long-wavelength sound modes and bulk flows in an s-wave superfluid. We have explicitly checked that when this formalism is applied to phonons it is equivalent to the usual approach based on a $P(X)$ effective field theory. We also considered phonon-roton scattering, and corrected earlier result by Landau and Khalatnikov as well as subsequent calculations based on similar techniques~\cite{khalatnikov1983phonon,um1988thermal}. Since an alternative EFT approach to rotons is presently not available, it is our hope that experiments will soon be able to weigh in on this discrepancy, given that interactions between phonons and rotons have already been the subject of very interesting experimental work~\cite{narayanamurti1976roton,forbes1990direct,faak2012roton}. Possible further applications of our formalism to liquid helium include analyses of phonon-mediated interactions between rotons~\cite{ishikawa1974phonon}, of roton-roton bound states \cite{greytak1970experimental}, of phonon emission by fast rotons  \cite{jackle1973emission}, and of light dark matter detectability \cite{Schutz:2016tid}.

We have also derived how a generic particle living in a superfluid couples to gravity. Although the associated effects for liquid helium in the Earth's gravitational field are probably too weak to be observable, we wonder whether they could be made much bigger by simulating  a much stronger gravitational field by means of inertial forces, e.g.~with a centrifuge\footnote{One should spin the system without producing vortex lines. In principle this can be achieved by setting up a $1/r$ profile for  the superfluid velocity, with $r$ being the distance from the centrifuge axis. Clearly the superfluid should be kept away from the $r=0$ axis.}. There could also be interesting applications to neutron star physics, where phonons are expected to have interesting seismological consequences \cite{Bedaque:2013fja}, and gravity is obviously much stronger. 

More in general, there are interesting consequences due to the  fact that for a non-relativistic superfluid an external gravitational field only appears as a shift of the chemical potential,
\be
\mu/m \to \mu/m -  \Phi \; ,
\ee
and that the phonon field also enters the effective theory as a modulation of the local chemical potential. These are particularly evident for ordinary objects immersed in our superfluid, for which gravitational phenomena are more readily observable. 

For instance, for a given object in a superfluid at equilibrium in the Earth's gravitational field, one could use a dynamometer to measure the net buoyancy, and then let the object go and measure the associated acceleration. According to the discussion in sect.~\ref{gravity}, these two measurements yield the quantities $E'_0$ and $M_{\rm eff}$. However, if one now imagines expanding the object's action \eqref{Sobj} in $\pi$ perturbations, one immediately sees that precisely these two quantities determine how our object couples to $\pi$ at linear order and at low speeds. And, so, by the simple gravitational experiment just described one can straightforwardly predict how our object will be shaken by an incoming sound wave, or how, conversely, it will generate sound waves if shaken or if  its volume ``pulsates." In our approach these phenomena are sensitive, respectively, to the combinations $(M_{\rm eff}+E_0' )/ M_{\rm eff}$ and $M_{\rm eff} + E_0'$, and to the time-dependence of $E_0'$. One can easily check that, with a different parametrization, precisely the same combinations enter the final results of \cite{landau:1987bo}, \S 11 and \S 74, which were derived by  more standard (and more laborious) hydrodynamical equations + boundary conditions techniques.

We will publish soon a more general analysis of how particles and small objects interact with superfluids and normal fluids, including the effects associated with anisotropies and spin.


\section*{Acknowledgments}  

We would like to thank Eric Cornell, Angelo Esposito, Lam Hui, Duccio Pappadopulo, Dam Thanh Son and Gabriele Trevisan for useful discussions. A.N. and R.P. are supported respectively by the NASA grant NNX16AB27G and the US Department of Energy (HEP) award DE-SC0013528.
 
\appendix
 

\section{Roton stability} \label{appa}
 
Consider a roton at rest, that is one with momentum $\vec p = p_* \hat n$ and energy $E(p_*) = \Delta$. By energy conservation, it can only decay to a combination of excitations ($i=1, \dots, N$) with lower energies, that is, excitations that, in the spectrum of fig.~\ref{fig:1}, lie to the left of $\bar p$. Momentum conservation reads
\be
\vec p = \vec p_1 + \dots + \vec p_N \; ,
\ee
which in particular implies
\be \label{momentum}
p_* = |\vec p \,  | \le |\vec p_1| + \dots + |\vec p_N| \; .
\ee

Let's now denote by $c_r$ the roton's {\em phase} velocity $\Delta/p_*$. It is a fact about superfluid helium that all the  excitations to the left of $\bar p$ have a {\em strictly} larger phase velocity, 
\be
\frac{E_i}{|\vec p_i|} >  c_r \; , \qquad i= 1, \dots N \; . 
\ee
Using this in the conservation of energy, 
\be \label{energy}
\Delta = E_1 + \dots + E_N \; , 
\ee
we get
\be
c_r p_* > c_r |\vec p_1| + \dots + c_r |\vec p_ N| \; ,
\ee 
in clear contradiction with eq.~\eqref{momentum}. This proves that, purely because of kinematical reasons, rotons at rest cannot decay.
 
Consider now a roton that has a small but non-vanishing group velocity. Its momentum is $\vec p = (p_*+\delta p) \hat n$, with $|\delta p| \ll p_*$, and its energy is $E \simeq \Delta + \delta p^2/(2m_*)$. It cannot decay to excitations to the left of $\bar p$ for the same reasons as above (the modifications to $c_r$ due to a non-vanishing $\delta p$ are negligible), but, by energy conservation, it could decay to a lower energy roton with momentum $\vec p \, ' = (p_* + \delta p') \hat n'$ (with $|\delta p'|<|\delta p|$) plus a combination of soft phonons, with very low momenta compared to $p_*$. In other words, it could emit soft phonons and slow down.
Momentum conservation now implies
\be
| \delta  p \,  | \le | \delta  p '  | +  |\vec p_1| + \dots + |\vec p_N| \; ,
\ee
while energy conservation reads
\be
\frac{\delta p^2}{2m_*} \simeq \frac{\delta p' {}^2}{2m_*} + E_1 + \dots + E_N \; .
\ee
Using $E_i = c_s |\vec p_i|$ for each phonon, these equations can be rewritten as
\begin{align}
|\vec p_1| + \dots + |\vec p_N| &\ge ( | \delta  p \,  | - | \delta p \,'  |) \\
|\vec p_1| + \dots + |\vec p_N| & \simeq \frac{1}{2 m_* c_s} ( | \delta  p \,  | + | \delta  p \,'  |)( | \delta  p \,  | - | \delta  p \,'  |) \ll( | \delta  p \,  | - | \delta  p \,'  |) \; ,
\end{align}
in clear contradiction with each other (we used that $|\delta p| , |\delta p'| \ll m_* c_s \sim p_*$). We thus see that all excitations close to the roton minimum are absolutely stable.


\section{Generic point particle-soft phonon interactions}\label{generic}

It is interesting to notice that interactions between point-like objects and soft phonons can be described in full generality without the need to specify the form of the function $f$ that appears in the effective action \eqref{Spp}. For example, to describe soft phonon emission (Cherenkov radiation) all we need are the interactions that are linear in $\pi$,
\be
f_v \, (\vec \nabla \pi \cdot \hat v) + f_X \, \dot \pi \; ,
\ee
whereas to describe  soft phonon scattering we also need  those that are quadratic in $\pi$,
\be
\sfrac12 (f_{vv}-f_v/v) (\vec \nabla \pi \cdot \hat v)^2 + \sfrac12(2f_v/v-f_X) (\vec \nabla \pi)^2 + \sfrac12 f_{XX} \, \dot \pi^2
+f_{vX} \, \dot \pi (\vec \nabla \pi \cdot \hat v)
 \; ,
\ee
as well as those that are bilinear in $\delta \vec x$ and $\pi$,
\be
f_{vX} \, \dot \pi (\delta \xdot \cdot \hat v) + (f_{vv}-f_v/v) (\delta \xdot \cdot \hat v)(\vec \nabla \pi \cdot \hat v) + (f_v/v-f_X) (\delta \xdot \cdot \vec \nabla \pi) + f_v \, \delta \vec x \cdot \vec \nabla(\vec \nabla \pi \cdot \hat v) \; ,
\ee
where the subscripts on $f$ denote derivatives with respect to its arguments, evaluated on the background (constant $\vec v = v \, \hat v$ and $X=\mu/m$). For scattering we also need the $\delta \vec x$ propagator, which we can get from its quadratic action,
\be
\sfrac12(f_{vv} \, P_{\parallel}^{ij} + f_v/v \, P_{\perp}^{ij}) \, \delta \dot x^i \delta \dot x^j \; .
\ee
The Feynman propagator thus is
\be
\langle \delta x^i \delta x^j \rangle = \frac{i}{\omega^2+ i\epsilon} \Big(\frac{1}{f_{vv}} P_{\parallel}^{ij}+ \frac{v}{ f_v} P_{\perp}^{ij}\Big) \; .
\ee

Using these expressions, one can easily calculate  emission rates and scattering cross-sections following procedures identical to those of  sects.~\ref{scattering} and \ref{sec:roton-phonon scattering}.

\section{Non-relativistic limit and coupling to gravity} \label{appb}

In this Appendix we will derive the effective action (\ref{P(X) action}) for a non-relativistic superfluid~\cite{Greiter:1989qb} starting from its relativistic analog~\cite{Son:2002zn} and taking the formal  $c \to \infty$ limit. This limit was also discussed in Appendix D of~\cite{Horn:2015zna}. Here, however, we will implement it in a slightly different way, taking into account also the coupling with a gravitational field. At the relativistic level, this coupling is achieved by placing the superfluid in a curved spacetime. Hence, our starting point will be the action
\begin{eqnarray} \la{rel P(X)}
S = \int d t d^3 x \sqrt{-g} \, P(X_{\rm r}), \qquad \quad X_{\rm r} = \sqrt{ - g^{\mu\nu} \d_\mu \phi \d_\nu \phi}, \qquad \quad \phi = \mu_{\rm r} t + \pi \; ,
\end{eqnarray}
where the subscript `r' stands for `relativistic'.

In order to make the non-relativistic limit more transparent, it is helpful to reintroduce all factors of $c$ explicitly. When $c \to \infty$, we only need to keep track of the perturbation to the $(0,0)$ component of the metric around Minkowski, which is related to the Newtonian potential by $ \delta g_{00} = - 2 \Phi /c^2$. It is also convenient to define the non-relativistic chemical potential $\mu \equiv \mu_{\rm r} - m c^2$, where $m$ is the mass of a helium atom. If we keep $\phi$ dimensionless as in \eqref{rel P(X)} and define $X_{\rm r} = c \sqrt{ - g^{\mu\nu} \d_\mu \phi \d_\nu \phi}$ so that it has units of energy, then $X_{\rm r}$ admits the following well-defined non-relativistic limit:
\begin{eqnarray} \la{NR limit of X}
X_{\rm r} = \sqrt{(1 - 2 \Phi /c^2)(mc^2 + \mu+ \dot \pi)^2 - c^2 (\nabla \pi)^2} \,\, \stackrel{c \to \infty}{\longrightarrow} \,\, mc^2 + \mu+ \left( \dot{ \pi} - \fr{(\nabla  \pi)^2}{2 m} - m \Phi \right) \! .
\end{eqnarray}
Thus, to lowest order in the derivatives the non-relativistic phonon field $\pi$ and the Newtonian potential $\Phi$ must always appear in the combination shown in parentheses, which can be thought of as a local modulation of the chemical potential $\mu$~\cite{Son:2005rv,Berezhiani:2015bqa}. We can also eliminate $m$---a microphysics quantity, not directly accessible from hydrodynamical experiments---from this combination of $\pi$ and $\Phi$, by rescaling the phonon field as $ \pi \equiv m \tilde \pi $. The non-relativistic action then depends only on the combination
\begin{eqnarray}
\delta X \equiv \dot{\tilde \pi} - \fr{(\nabla \tilde \pi)^2}{2 } -  \Phi,
\end{eqnarray}
which is independent of $m$ and reduces to the result in eq. (\ref{P(X) action}) once the Newtonian potential is turned off (and tildas are dropped)\footnote{Note that with our conventions $X_{\rm r}$ and $\delta X$ have different units: the former has the dimensions of an energy, while the latter has dimensions of a velocity squared, that is, energy per unit mass. Similarly, $\pi$ is dimensionless whereas $\tilde \pi$ has dimensions of an inverse mass.
}. 
Notice however that $m$ reappears now in  how $\delta X$ affects the local chemical potential, $\mu + m \, \delta X$. 
This is because $m$ is the proportionality factor between mass density and number (or charge) density \cite{Greiter:1989qb}, and the normalization of the chemical potential knows about that of the charge. We could decide to never talk about the number of particles, and only about the total mass of the system, which, unlike the number, is directly accessible by macroscopic measurements. In that case the associated chemical potential would have units of energy per unit mass, like $\delta X$,  and $m$ would not appear anywhere in the effective theory. This is probably the best choice of normalizations from an effective field theory standpoint, but in fact, to make referring to experimental data more transparent, we will stick to the usual definition of the chemical potential.

The RHS of eq. (\ref{NR limit of X}) shows that in order to couple a non-relativis	tic superfluid to gravity it is sufficient to shift the non-relativistic chemical potential: $\mu \to \mu -m \Phi$. Note in fact that the determinant of the metric in eq. (\ref{rel P(X)}) becomes trivial in the $c \to \infty$ limit. For a superfluid at equlibrium, it is easy to understand the origin of this shift using a thermodynamic argument. Let's consider a region of space that contains a macroscopic number $N$ of helium atoms, and yet that is small enough so that the  external gravitational potential $\Phi$ is approximately constant. Each atom in this region acquires a potential energy $ m \Phi$. As a result, the total Gibbs free energy at pressure $\bar P$ becomes $G (\bar P) = G_0 (\bar P)+ N m \Phi$, where $G_0$ is the Gibbs free energy one would have at the same pressure but in the absence of gravity (we are setting the temperature to zero, which is the case we are interested in). The total chemical potential then is given by:
\begin{eqnarray} \label{chemical potential with gravity Landau}
\mu = \l. \fr{\d G}{\d N} \r|_{\bar P} = \mu_0 (\bar P ) + m \Phi,
\end{eqnarray}
where $\mu_0$ is the chemical potential one would have without gravity. Thermodynamic equilibrium requires $\mu$ to be constant throughout the system~\cite{Landau:1980mil}.

The same result follows immediately from our EFT approach. For any given  pressure $\bar P$, static equilibrium requires $\pi = \text{constant}$ and so
\begin{eqnarray}
\bar P = P( \mu - m \Phi ) \; ,
\end{eqnarray}
since $P(X)$ is nothing but the pressure (eq. (\ref{p rho u}) holds for a curved spacetime as well). Then, we can invert the relation above to find
\begin{eqnarray}
\mu = P^{-1} (\bar P) + m \Phi \; ,
\end{eqnarray}
which is precisely of the form (\ref{chemical potential with gravity Landau}), since $P^{-1} (\bar P)$ is the chemical potential we would have at pressure $\bar P$ if $\Phi $ were zero. 
In the next Appendix, we will see that in fact the replacement $\mu \to \mu - m \Phi$ is also the correct prescription to couple point-like particles such as phonons and rotons to gravity.

\section{Effective action from the coset construction} \la{appc}

We will now give an alternative derivation of the effective action for a point-like object moving in a superfluid. Our discussion will be entirely based on symmetry considerations and rely on the coset construction~\cite{Coleman:1969sm,Callan:1969sn} for spontaneously broken space-time symmetries~\cite{Volkov:1973vd,ogievetsky:1974ab}. In what follows we will assume familiarity with this technique. If necessary, we refer the reader to Section 2 of~\cite{Delacretaz:2014oxa} for a concise review.

The symmetry group of a non-relativistic (s-wave\footnote{For p-wave superfluids one needs to consider  the spin as well, which, for non-relativistic systems with negligible spin-orbit couplings, can be treated as an internal symmetry.}) superfluid is simply the Galilei group. Its generators satisfy the well-known commutation relations~\cite{Weinberg:1995mt}
\begin{align} \la{galilei algebra}
&\small[ J_i, J_j\small] = i \epsilon_{ijk} J^k, \qquad\quad  \small[ J_i, K_j\small] = i \epsilon_{ijk} K^k, \qquad\quad  \small[ J_i, P_j\small] = i \epsilon_{ijk} P^k, \\
& \qquad \qquad \small[ K_i, P_j \small] = - i \delta_{ij} M, \qquad\quad \small[K_i, H_0 \small] = - i P_i. \nonumber
\end{align}
The algebra above contains an additional central charge, $M$, compared to the Poincar\'e algebra. This is just the total mass, which for non-relativistic systems is a conserved quantity. For systems made of a single particle species, $M$ is simply proportional to the particle number $Q$, i.e.~$M = m Q$, with $m$ the mass of the particle.\footnote{Note that if we trade $M$ for $Q$ in the algebra, the mass $m$ can always be removed by rescaling appropriately the generators $K_i$ and $P_i$. Since our derivation of the action for the Goldstone modes will solely depend on the symmetry algebra and the breaking pattern, this is  another manifestation of the fact that one doesn't need to know the mass of the elementary constituents (helium atoms, in our case) to describe their collective excitations (phonons).}
A superfluid state spontaneously breaks some of the Galilei symmetries in a way that can be summarized as follows~\cite{Endlich:2013spa}:
\begin{eqnarray} \la{superfluid SBP}
\text{unbroken} = \l\{ \begin{array}{l} H \equiv H_0 - \mu Q \\ \vec P \\ \vec J \end{array} \r. \qquad \quad \text{broken} = \l\{ \begin{array}{l} Q \\ \vec K \end{array} \r.
\end{eqnarray}
Boosts are broken because the superfluid admits a preferred reference frame---the one in which it is at rest---and so is $Q$ (this is the field theory analog of Bose-Einstein condensation)~\cite{Endlich:2013spa}. On the other hand, the ground state is still homogeneous and isotropic, which is why $\vec P$ and $\vec J$ are unbroken, and it is an eigenstate of the combination $H_0 - \mu Q$---the effective Hamiltonian at finite chemical potential---which therefore is also unbroken.

Using the coset construction, one can derive the effective action (\ref{P(X) action}) for a superfluid solely from the Galilei algebra (\ref{galilei algebra}) and the symmetry breaking pattern (\ref{superfluid SBP}). Let's review very briefly how this works in this simpler setting before turning our attention to the more involved point-like particle case. A more pedagogical derivation of what follows can be found in~\cite{Endlich:2013spa}. Our starting point is the coset parametrization
\begin{eqnarray}
\Omega = e^{- i t H + i \vec x \cdot \vec P} e^{i \vec \eta (t, \vec x)\cdot \vec K } e^{i \pi  (t, \vec x) Q},
\end{eqnarray}
out of which one can build the Maurer-Cartan form:
\begin{eqnarray} \la{MC form superfluid}
\Omega^{-1} \d_\mu \Omega = i \l\{ - \delta_\mu^0 H + (\delta_\mu^i + \delta_\mu^0 \eta^i) P_i + \d_\mu \eta^i K_i + \l(\d_\mu \pi - \delta_\mu^i m \eta_i - \delta_\mu^0 \fr{m}{2} \eta^2 \r) Q\r\} ,
\end{eqnarray}
which can be easily obtained by repeated use of the commutation relations (\ref{galilei algebra}). The fields $\pi$ and $\vec \eta$ are the Goldstone modes associated with the breaking of $Q$ and $\vec K$ respectively, and their ``covariant derivatives'' can be extracted~\cite{Endlich:2013spa} from the RHS of (\ref{MC form superfluid}):
\begin{eqnarray}
\nabla_\mu \pi &=& \delta_\mu^0 \l(\dot \pi +\fr{m}{2} \eta^2 - \eta^i \d_i \pi \r) + \delta_\mu^i \l( \d_i \pi - m  \eta_i \r) \\
\nabla_\mu \vec \eta &=& \d_\mu \vec \eta.
\end{eqnarray}
It is well known that s-wave superfluids at $T=0$ have only one Goldstone mode---the superfluid phonon. In fact, the Goldstones $\vec \eta$ can be eliminated from the effective theory in a way that is consistent with all the symmetries by imposing a so-called inverse Higgs constraint~\cite{Ivanov:1975zq}. In our case, this amounts to setting to zero the spatial covariant derivative of $\pi$ and solving for $\vec \eta$:
\begin{eqnarray} \la{inverse Higgs constraint superfluid}
\nabla_i \pi = \d_i \pi - m \eta_i \equiv 0 \qquad \longrightarrow \qquad \eta_i = \fr{\d_i \pi}{m}.
\end{eqnarray}
Then, the only quantity that remains with at most one derivative per field is
\begin{eqnarray}
\nabla_t \pi = \dot \pi +\fr{m}{2} \eta^2 - \eta^i \d_i \pi = \d_t \pi - \fr{(\nabla \pi)^2}{2m}.
\end{eqnarray}
which up to an inconsequantial factor of $m$ is precisely the combination that appears in eq.~(\ref{P(X) action}).

In the Newtonian limit, the coupling with gravity is obtained by replacing $\d_\mu \to \d_\mu - i \delta_\mu^0 \Phi M$ in the definition of the Maurer-Cartan form~\cite{Delacretaz:2014oxa,Brauner:2014jaa}. This replacement has the only effect of changing the covariant derivative of $\pi$ to:
\begin{eqnarray} \la{IH constraint superfluid}
\nabla_\mu \pi = \delta_\mu^0 \l(\dot \pi +\fr{m}{2} \eta^2 - \eta^i \d_i \pi - m \Phi\r) + \delta_\mu^i \l( \d_i \pi - m  \eta_i \r).
\end{eqnarray}
Thus, the inverse Higgs constraint remains the same as in eq. (\ref{inverse Higgs constraint superfluid}), and after expressing $\nabla_t \pi$ solely in terms of $\pi$ we recover the combination in parentheses on the RHS of (\ref{NR limit of X}).

Let us now turn our attention to the case of a point-like object moving in the superfluid with some constant velocity $\vec v_0 = v_0 \hat x_3$, and let us determine how the action can depend on perturbations of its trajectory. We take this as a starting point, rather than the case of an object at rest, because it is more general: for instance, phonons can never be at rest.
The object in question will break additional symmetries compared to the ones already broken by the medium. In fact, the overall system is  no longer homogeneous, since we can think of the object as some sort of impurity, nor isotropic, since the motion of the object defines a preferred direction. The only symmetries that remain unbroken are therefore
\begin{eqnarray}
H' \equiv H - v_0 P_3, \qquad \qquad J_3 \; .
\end{eqnarray}
All other symmetries are spontaneously broken, and thus the coset parametrization is now more involved and reads\footnote{The actions for relativistic and non-relativistic point-like particles in vacuum were derived using the coset construction in~\cite{Delacretaz:2014oxa,McArthur:2010zm} and~\cite{Goon:2012dy} respectively.}
\begin{eqnarray}
\Omega(t) = e^{- i t H'}e^{ i \delta \vec x (t) \cdot \vec P} e^{i \vec \eta (t, \vec x(t))\cdot \vec K } e^{i \pi (t, \vec x(t)) Q} e^{i \theta^a (t) J_a},
\end{eqnarray}
with $J_a = (J_1, J_2)$.\footnote{Notice that our system will in general feature more than one symmetry breaking scale. In fact, $Q$ and $\vec K$ (\ref{superfluid SBP}) are broken by the medium at length scales of the order of the interatomic distance $a$; $\vec P$ is instead broken by the object at a length scale of the order of its size $R$, which for phonons and rotons is determined by the de Broglie wavelength, $1/p$; finally, $J^a$ is broken at a time scale of the order of $R /v_0$, since this is the time scale over which the motion of the object becomes observable. This time scale can be converted into a length scale using the smallest available sound speed, which can be either that of the superfluid or that of the material that makes up the object.} Here, $ \delta \vec x(t)$ should be thought of as fluctuations in the position of the object around the background solution $\vec x_0 (t) = \vec v_0 t$. Similarly, $\theta^a (t)$ should be interpreted as fluctuations of the direction of the particle's trajectory. From this perspective, the $\theta$'s are in some sense redundant, and in fact we will be able eliminate them by imposing more inverse Higgs constraints. Notice also that bulk fields such as $\pi$ and $\vec \eta$ must evaluated at the instantaneous position of the object, that is $\vec x(t) = \vec v_0 t + \delta \vec x (t)$. The Maurer-Cartan form including the coupling with gravity is now
\begin{eqnarray}
\Omega^{-1} \Big[\frac d{dt} - i  \Phi (t, \vec x(t) )M \Big] \Omega
\end{eqnarray}
and the relevant covariant derivatives are
\begin{eqnarray}
\nabla_t x^i &=& (\dot x^j + \eta^j)R_j{}^i (\theta^a)- v_0 \delta_3^i\\
\nabla_t \pi &=& \dot \pi - \fr{m}{2} \eta^2 - m \Phi + \dot{\vec x} \cdot (\nabla \pi - m \vec \eta).
\end{eqnarray}
with $R_{ij} = \l(e^{i \theta^a J_a}\r)_{ij}$. These covariant derivatives have only a time component because the only unbroken translations are those generated by $H'$.

Using the solution to the bulk inverse Higgs constraint (\ref{inverse Higgs constraint superfluid}), we see immediately that $\nabla_t \pi$ reduces to the combination of bulk fields $\pi$ and $\Phi$ that appears on the RHS of eq. (\ref{NR limit of X}). Moreover, as long as $v_0 \neq 0$, we can also impose the inverse Higgs constraints
\begin{eqnarray} \la{IH theta}
\nabla_t x^a = (\dot x^j + \eta^j)R_j{}^a \equiv 0,
\end{eqnarray}
and solve them to eliminate the Goldstones $\theta_a$. The easiest way to achieve this is to realize that, thanks to the properties of rotation matrices, we can regard $R_j{}^1, R_j{}^2$ and $R_j{}^3$ as three othonormal vectors. From this viewpoint, the inverse Higgs constraint (\ref{IH theta}) implies that $(\dot x_j + \eta_j) \propto R_j{}^3$, and requiring that $R_j{}^3 R^{j3} = 1$ we find
\begin{eqnarray} \la{nabla_t x^3}
\nabla_t x^3 = | \dot {\vec x} + \nabla \pi / m| - v_0.
\end{eqnarray}
Since the superfluid velocity in these units is $\vec u = - \nabla \pi /m$, we thus see that the low-energy effective action for a point-particle moving  in a superfluid is
\begin{eqnarray} \label{pp with gravity}
S_{\rm p.p.} = \int dt f (\dot \pi - (\nabla \pi)^2 /2m -m \Phi, | \dot {\vec x} - \vec u| ),
\end{eqnarray}
in complete agreement with the more heuristic arguments given in Sec. \ref{effective}.

We should point out that the quantity (\ref{nabla_t x^3}) admits a well-defined expansion in powers of the Goldstones $\delta \vec x$ and $\pi$ only because we are expanding around a non-trivial background $\dot {\vec x} = \vec v_0$. Conversely, the constraint (\ref{IH theta}) doesn't admit a local solution for $\theta^a$ when $v_0 =0$. This statement is equivalent to the observation made in Sec. \ref{effective} that the variable $\hat p$ cannot be integrated out for rotons with zero velocity.

To conclude, notice that there is a subtlety similar to the one discussed around eq.~\eqref{v^2}. Like in that case, we could add to the point-particle action a term of the form
\be \label{M Phi}
-\int dt \, M_g \Phi \; ,
\ee
which is in fact the correct coupling to Newtonian gravity for a particle in empty space, $M_g$ being its gravitational mass. This  term is invariant under the Newtonian gauge transformations $\Phi \to \Phi + {\rm const}$ only up to total derivatives, which is why it's not contemplated by the structure  \eqref{pp with gravity}. However, from general relativity we know that $M_g$ must be the same as the $M_i$ appearing in \eqref{v^2}---both terms come from the non-relativistic limit of $- M_i \int dt \sqrt{g_{\mu\nu} \, \dot x ^\mu \dot x^\nu}$. And so,  following exactly the same manipulations as in \eqref{manipulate}, the sum of \eqref{v^2} and \eqref{M Phi} can be completely reabsorbed into the structure \eqref{pp with gravity}.
 

\bibliography{biblio}{}
\bibliographystyle{hieeetr}

\end{fmffile}

\end{document}